\documentclass[structabstract]{aa}

\usepackage{txfonts}
\usepackage{graphicx}
\usepackage{natbib}
\usepackage{mathrsfs}
\usepackage{rotating}
\usepackage{color}  

\def\lso{L_\odot}
\def\llso{\log\,L/L_\odot}
\def\mso{M_\odot}
\def\msoy{\mso \, {\rm yr}^{-1}}
\def\Teff{T_{\rm eff}}

\def\simgr{\,\hbox{\hbox{$ > $}\kern -0.8em \lower 1.0ex\hbox{$\sim$}}\,}
\def\simle{\,\hbox{\hbox{$ < $}\kern -0.8em \lower 1.0ex\hbox{$\sim$}}\,}

\bibpunct{(}{)}{;}{a}{}{,}

\title{The evolution of rotating very massive stars with LMC composition \footnote{The dataset of the presented stellar evolution models is available in electronic form at the CDS via anonymous ftp to cdsarc.u-strasbg.fr (130.79.128.5) or via http://cdsweb.u-strasbg.fr/cgi-bin/qcat?J/A+A/}}

\author{K. K\"ohler\inst{1},
N. Langer\inst{1},
A. de Koter\inst{2,3},
S.E. de Mink\inst{2,4,5},
P.A. Crowther\inst{6},
C.J. Evans\inst{7},
G. Gr\"afener\inst{8},
H. Sana\inst{9},
D. Sanyal\inst{1},
F.R.N. Schneider\inst{1},
J.S. Vink\inst{8}}

\institute{Argelander-Institut f\"ur Astronomie der Universit\"at Bonn, Auf dem H\"ugel 71, 53121 Bonn, Germany
\and
Astronomical Institute ``Anton Pannekoek'', Amsterdam University, Science Park 904, 1098 XH, Amsterdam, The Netherlands
\and
Instituut voor Sterrenkunde, KU Leuven, Celestijnenlaan 200D, 3011 Leuven, Belgium
\and
Observatories of the Carnegie Institution for Science, 813 Santa Barbara St, Pasadena, CA 91101, USA
\and
Cahill Center for Astrophysics, California Institute of Technology, Pasadena, CA 91125, USA
\and
Department of Physics \& Astronomy, University of Sheffield, Sheffield, S3 7RH
\and
UK Astronomy Technology Centre, Royal Observatory Edinburgh, Blackford Hill, Edinburgh, EH9 3HJ, UK
\and
Armagh Observatory, College Hill, Armagh BT61 9DG, UK
\and
Space Telescope Science Institute, 3700 San Martin Drive, Baltimore, MD 21218, USA
}

\date{Received date / Accepted date}

\abstract 
{With growing evidence for the existence of very massive stars at subsolar metallicity,
there is an increased need for corresponding stellar evolution models.}
{We present a dense model grid with a tailored input chemical composition
appropriate for the Large Magellanic Cloud.}
{We use a one-dimensional hydrodynamic stellar evolution code, which accounts for rotation, transport of angular momentum by magnetic fields, and stellar wind mass loss to compute our detailed models. We calculate stellar evolution models with initial masses from 70 to 500\,$M_{\odot}$ and with initial surface rotational velocities from 0 to 550\,km/s, covering the core-hydrogen burning phase of evolution. }
{We find our rapid rotators to be strongly influenced by rotationally induced mixing of helium, with quasi-chemically homogeneous evolution occurring for the fastest rotating models. Above $160\mso$, homogeneous evolution is also established through mass loss, producing pure helium stars at core hydrogen 
exhaustion independent of the initial rotation rate. Surface nitrogen enrichment is also found for slower rotators, even for stars that lose only a small fraction of their initial mass.
For models above $\sim 150\mso$ at zero age, and for models in the whole considered mass range later on, we find a considerable envelope inflation due to the proximity of these models to their Eddington limit. This leads to a maximum zero-age main sequence surface temperature of $\sim 56\,000$\,K, at $\sim 180\,M_{\odot}$, and to an evolution of stars in the mass range $50\mso\dots 100\mso$ to the regime of luminous blue variables in the HR [spell out] diagram with high internal Eddington factors. Inflation also leads to decreasing surface temperatures during the chemically homogeneous evolution of stars above $\sim 180\,M_{\odot}$.}
{The cool surface temperatures due to the envelope inflation in our models lead to an enhanced mass loss, which prevents stars at LMC metallicity from evolving into pair-instability supernovae. The corresponding spin-down will also prevent very massive LMC stars to produce long-duration gamma-ray bursts, which might, however, originate from lower masses.}


\usepackage{color}
\begin{document}

\titlerunning{Evolution of rotating very massive stars}
\authorrunning{K. K\"ohler et al.}
\maketitle

\section{Introduction}
\label{sec1}
Massive stars, with initial masses above $\sim 8\mso$, are powerful cosmic engines \citep{Bresolin_2008}. They produce
copious amounts of ionizing photons, strong stellar winds, energetic final explosions, and
most of the heavy elements in the Universe.
The most massive amongst them conduct a life very close to their Eddington limit, and may thus be prone
to become unstable. They are thought to be able to produce the most spectacular stellar explosions,
like hypernovae, pair-instability supernovae, and long-duration gamma-ray bursts \citep{Langer_2012}.

While the value of the upper mass limit of stars is presently uncertain 
\citep{Schneider_2014}, there is a great deal of evidence of stars with initial masses well
above $100\mso$ in the local Universe. 
A number of close binary stars have been found with component initial masses above $100\mso$
\citep{Schnurr_2008,Schnurr_2009,Sana_2013}. 
\citet{Crowther_2010}
proposed initial masses of up to $300\mso$ for several stars in the Large Magellanic Cloud (LMC)
based on their luminosities.

We present stellar evolution models of very massive rotating core-hydrogen burning
stars, with initial masses 
up to 500\,$M_{\odot}$. These models will be used for comparison with the VLT FLAMES 
Tarantula survey \citep{Evans_2011},
which observed more than 800 O~and early-B stars
in the 30 Doradus region located in the Large Magellanic Cloud, to study the effects 
of rotational mixing and mass loss on the evolution of very massive stars. 

Stellar evolution models for very massive stars have already been 
presented (see Table~1 in \citet{Maeder_2011} and 
references therein). In particular, we refer to the rotating stellar 
models of \citet{Crowther_2010,Yusof_2013},  who calculated models with initial masses up to 500\,$M_{\odot}$. 
It is the aim of this paper to present a grid of stellar evolution models that has a dense spacing in mass and 
rotation rates, such that it is suitable for forthcoming population synthesis calculations. 
In this sense, our models are an extension of the LMC models of \citet{Brott_2011} to higher masses.
In the mass range considered here, stellar wind mass loss and the proximity to the Eddington limit
play prominent roles.



In Sect.~\ref{sec2} we briefly describe the employed stellar evolution code, including the input parameters for 
our calculations. 
Following that, we present and discuss the stellar evolution 
models in Sect.~\ref{sec3}, including the evolution of their mass and their surface abundances, with emphasis on 
chemically homogeneous evolution, and we compare our models with previous work in Sect.~\ref{sec4}. Section~\ref{sec5} contains our short summary. In Appendix~\ref{app1}, we present isochrones derived from our models, as functions
of age and initial stellar rotation rate.

\section{Input physics and assumptions}
\label{sec2}

\citet{Brott_2011} computed three grids of stellar evolution models for
different metallicities (Galactic, SMC, LMC) for initial masses up to
60\,$M_{\odot}$, which were compared with the results of 
the VLT-FLAMES Survey of Massive Stars \citep{Evans_2005,Evans_2006}.
In particular, the convective core overshooting parameter and the efficiency
parameters for rotationally induced mixing used by \citet{Brott_2011} were
calibrated to reproduce the results of this survey \citep{Hunter_2008}.
The dense spacing in initial mass and rotational velocity used for the grid
of \citet{Brott_2011} opened the door for statistical tests of the theory of 
massive star evolution (\citet{Brott_2011b}; Schneider et al., in preparation).
Here, we extend the LMC grid of \citeauthor{Brott_2011} up to initial masses of 500\,$M_{\odot}$,
using the same numerical code, physical assumptions, and a similarly dense grid spacing.

To calculate rotating stellar evolution models, we use our one-dimensional hydrodynamic binary stellar evolution code (BEC) 
which is described in \citet{Heger_2000,Petrovic_2005}, and \citet{Yoon_2012}. 
It contains a detailed treatment of rotation, 
angular momentum transport due to internal magnetic fields, and stellar wind mass loss. 
The code solves all five stellar structure equations throughout the stellar interior,
including the stellar envelope up to a Rosseland optical depth of $\tau=2/3$. Convection is 
considered throughout the star using the non-adiabatic mixing-length theory.
Our code is suited to treating stars close to the Eddington limit, and to describe
the effects of envelope inflation which occur in this situation \citep{Ishii_1999,Petrovic_2006}.

\subsection{Chemical composition}

The initial chemical composition for our models is chosen according to corresponding observations 
of young massive stars and of H\,{\sc ii}-regions in the LMC \citep{Brott_2011}. We thus adopt 
initial mass fractions for hydrogen, helium, and the sum of all metals of 
$X=0.7391$, $Y=0.2562,$ and $Z=0.0047$, respectively, with a non-solar metal abundance pattern
(see Tables~1 and~2 in \citeauthor{Brott_2011}). We note that the applied opacities and mass loss rates (see below)
are scaled with the LMC iron abundance, not with the total metallicity, which is 
reduced by 0.35\,dex with respect to that of the Sun. 

\subsection{Convection and rotational mixing}

Convection with a mixing-length parameter of $\alpha_{\mathrm{MLT}} = 1.5$ \citep{Vitense_1958,Langer_1991} is 
applied as well as semi-convection with an efficiency parameter of $\alpha_{\mathrm{SEM}} = 1$ \citep{Langer_1983,Langer_1991}. 
In addition to convection, convective core overshooting is included with $\alpha_{\rm over}=0.335$ local pressure scale heights, 
as calibrated in \citet{Brott_2011} with the rotational properties of B-type stars \citep{Hunter_2008,Vink_2010} from the 
VLT-FLAMES survey. While no observational calibration of the overshooting parameter exists for stars of the considered mass range,
we point out that the role of overshooting in our models is  minor because of the large convective core mass fractions
of very massive stars.

Rotational mixing \citep{Heger_2000} is considered with the efficiency parameters of $f_{\rm c} = 0.0228$ 
and $f_{\mu}=0.1$ \citep{Brott_2011}. 
The most significant process causing rotationally induced mixing in our models is the Eddington Sweet circulation.
Furthermore, the transport of angular momentum by magnetic fields due to the Spruit-Taylor 
dynamo \citep{Spruit_2002} is applied, which is assumed here not to lead to an additional
transport of chemical elements \citep{Spruit_2006}.
Since the magnetic torques lead to a shallow angular velocity profile in our models, the effects of the shear instability, although included, are quite limited during the main sequence evolution.

\subsection{Mass loss}

The evolution of very massive stars is intimately connected to their mass-loss behaviour. 
Mass loss of very massive stars, for which few observational constraints exist, is a 
very active field of research.  Here, we provide a brief overview of the mass loss 
prescriptions adopted in our calculations.

From the zero-age main sequence up to a surface helium fraction $Y_{\rm s} = 0.4$, we use the mass-loss predictions by
\citet{Vink_2000,Vink_2001} for O and B stars.  These rates are valid for stars of 
one million solar luminosities or less, i.e. stars below $80\,\mso$ \citep{Mokiem_2007}, which are
not very close to their Eddington limit (i.e. have $\Gamma \la 0.3$; see Eq.~\ref{eq:Gamma}). Empirical
tests of these predictions depend critically on the presence and properties of small scale structures in the
outflows, known as clumping.  \citet{Mokiem_2007} showed on the basis of H$\alpha$ and He~{\sc ii}\,$\lambda$4686
analysis that the predicted rates agree with observations if the material is concentrated in clumps that have a
3--4 times higher density than in a smooth outflow.  Other analyses, which include modelling of ultraviolet resonance lines, derive clumping factors that may reach values of~10 (e.g. \citet{Bouret_2004,Bouret_2005,Bouret_2012,Fullerton_2006}).  In this last case, the Vink et al. prescription adopted here may overestimate 
$\dot{M}$ by about a factor of 2.  An improved hydrodynamical treatment shows that for normal
O stars the mass loss rates may be somewhat lower \citep{Mueller_2008,Muijres_2012}.

Predictions for stars in the 40--300\,$\mso$ range have been presented by \citet{Vink_2011}. Objects at the
upper mass end of this range may have a very high Eddington factor $\Gamma$. It has been found that, at solar metallicity, 
the wind strengths agree with the standard Vink et al. recipe for objects that have an Eddington factor for Thomson 
scattering of $\Gamma_{\rm e} \la 0.7$.  
For higher $\Gamma_{\rm e}$ values these new predictions show an upturn in the mass-loss rate, leading to rates 
that are higher by up to about a factor of 2 compared to the Vink et al. values used here.  The authors associate
the upturn with the stellar wind becoming optically thick, i.e. leading to spectral morphology which is typical for Wolf-Rayet stars of nitrogen subclass (WN) showing hydrogen emission lines. Interestingly, this predicted upturn may have been confirmed observationally \citep{Bestenlehner_2014}.  
The relation between mass loss and Eddington factor has also been explored for late-WN stars by 
\citet{Graefener_2008} and \citet{Graefener_2011}. They find a behaviour that is
similar to the results of \citet{Vink_2011}, though the onset of Wolf-Rayet type outflows occurs at lower 
$\Gamma_{\rm e}$ values. \citeauthor{Graefener_2008} however report a temperature dependence of their mass-loss
rates that is steeper than that of \citet{Vink_2011}.

Since the mass loss predictions for large Eddington factors cannot yet be implemented unambiguously into stellar
evolution calculations, we  extrapolate the \citet{Vink_2001} rates for stars above $10^6\lso$. We note that
\citet{Crowther_2010} found the Vink et al. rates to agree within error bars with those observed in stars
of up to $\sim 10^7\lso$ found in 30~Doradus. 
For objects in the range 60$\dots$100$\,\mso$ the objects are close to the model-independent mass-loss transition 
point between optically thin and thick winds. In this range mass-loss rates have recently been calibrated 
with an uncertainty of only $\sim$30\% \citep{Vink_2012}.

The Vink et al. mass-loss prescription shows a bi-stability jump at about 25\,000\,K, leading to an increase of the mass-loss rate by a factor of 5 for stars of spectral type B1.5 or later.  We include this bi-stability jump in our
calculations (cf. \citet{Brott_2011}). 
%
Additionally the \citet{Nieuwenhuijzen_1990} empirical 
mass-loss rate is applied to cope with an increase in mass loss when approaching the Humphreys-Davidson limit (HD limit). 
The transition from \citet{Vink_2000,Vink_2001} to \citet{Nieuwenhuijzen_1990} occurs at any effective temperature smaller 
than 22\,000 K where the \citet{Nieuwenhuijzen_1990} mass-loss rate exceeds the \citet{Vink_2000,Vink_2001} mass-loss 
rate. 

To account for Wolf-Rayet mass loss, the \citet{Hamann_1995} mass-loss rate divided by a factor of 10 
is applied for the surface helium fraction $Y_s \ge 0.7$. 
Figure~1 in \citet{Yoon_2010} shows that this corresponds well to the 
Wolf-Rayet mass-loss rate proposed by \citet{Nugis_2000} for Wolf-Rayet
masses in the range 5$\mso$ to $20\mso$. 
For surface helium mass fractions from $0.4 \le Y_s \le 0.7$ 
the mass-loss rate is linearly interpolated between either the \citet{Vink_2000,Vink_2001} or the 
\citet{Nieuwenhuijzen_1990} mass-loss rate and that of \citet{Hamann_1995}.

We apply a mass-loss enhancement for stars near critical rotation as in \citet{Yoon_2005}
which considers a reduction of the critical rotational velocity for stars near their Eddington limit.
It is still unclear whether rapid rotation per se leads to an enhanced mass loss \citep{Mueller_2014}, but it appears reasonable to consider that
the mass-loss rate increases close to the Eddington limit, which is indeed reached sooner for rotating objects
\citep{Langer_1997}.

\begin{figure}[htbp]
    \centering
     \includegraphics[angle=-90,width=8.5cm]{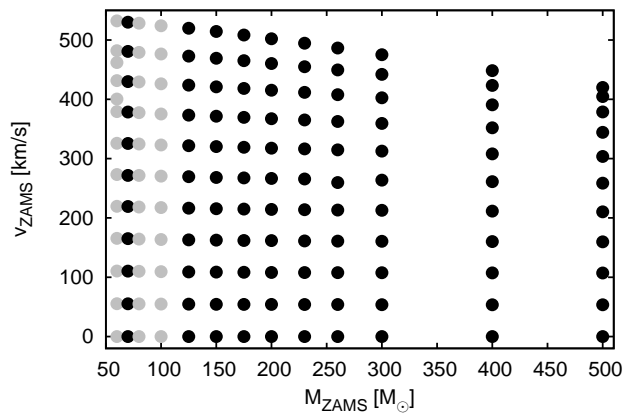}
    \caption{Initial equatorial rotational velocity versus initial mass. Each dot in this diagram represents the evolutionary sequence in our model grid with the corresponding initial parameters. 
Grey dots correspond to models presented in \citet{Brott_2011} as well as previously unpublished models
calculated by I. Brott, while black dots represent the 110 newly computed evolutionary sequences. }
    \label{fig_grid}
\end{figure}

\subsection{Model grid}

Figure~\ref{fig_grid} gives an overview of our grid of evolution models
by indicating the initial masses and initial surface rotational velocities of all computed model sequences.
Because of the increase in the Eddington factor with mass, we decrease the maximum initial
rotational velocity for higher masses in order to avoid strong rotationally induced mass loss already on the
zero-age main sequence \citep{Langer_1998}. 
Whereas most sequences are computed to core-hydrogen exhaustion, some of the most massive and most rapidly rotating
models were stopped shortly before their proximity to the Eddington limit caused numerical difficulties.

\section{Results}
\label{sec3}
\subsection{Evolutionary tracks in the HR-diagram}
\label{sec3_trac}

\begin{figure*}[htbp]
    \centering
     \includegraphics[angle=-90,width=18cm]{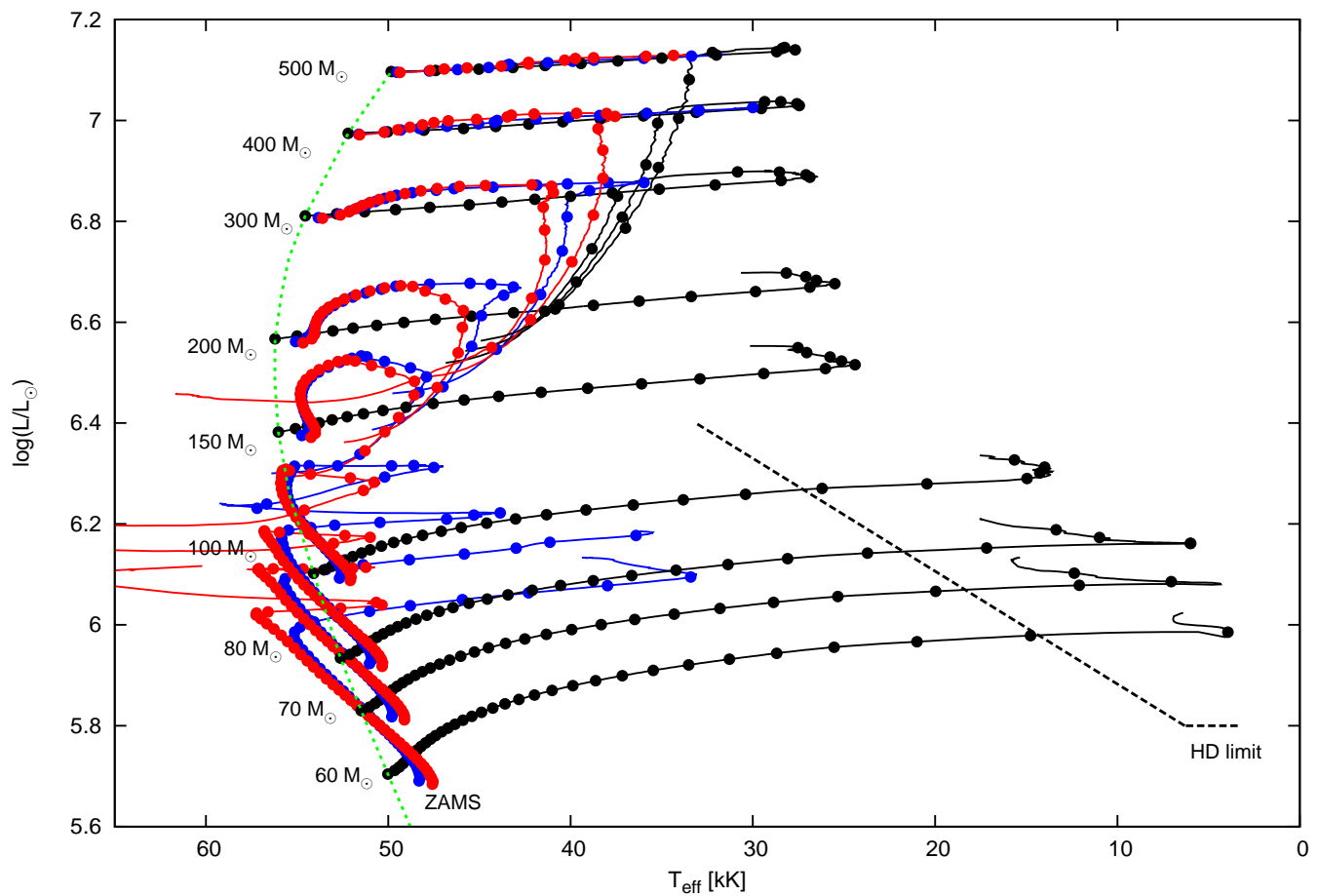}
    \caption{Evolutionary tracks of massive stars during their core hydrogen burning evolution in  
the Hertzsprung-Russell diagram. For each selected initial mass (as labelled), tracks are shown for three
different initial surface rotational velocities, $\varv_\mathrm{ZAMS}=0, 400, 500$\,km/s, in black, blue, and red, respectively.
The time difference between two successive dots on each track is 10$^5$ yr. The zero-age main sequence is 
drawn as a green dashed line. The end of the tracks corresponds to the terminal age main sequence.
The approximate location of the Humphreys-Davidson limit is indicated by the black dashed line \citep{Humphreys_1994}. 
}
    \label{fig_hrd}
\end{figure*}

A selection of stellar evolution tracks is presented in Fig.~\ref{fig_hrd}. The luminosity of the stellar 
models for a given initial mass and initial surface rotational velocity is shown as a function of their effective temperature. 
Tracks are shown for nine different initial masses from 60\,$M_{\odot}$ to 500\,$M_{\odot}$, with initial surface rotational 
velocities of 0\,km/s, 400\,km/s, and 500\,km/s.

        \begin{figure}[htbp]
    \centering
     \includegraphics[angle=-90,width=8cm]{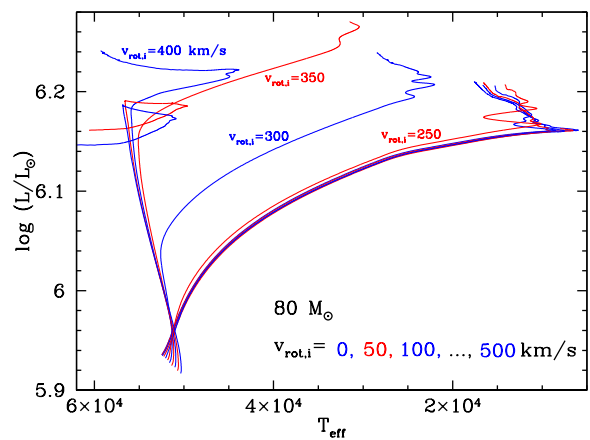}
    \caption{Evolutionary tracks of stars with an initial mass of 80$\mso$, for initial rotational velocities of
0, 50, 100, 150, 200, 250, 300, 350, 400, 450, and 500\,km/s. Tracks with initial velocities that are even
multiples of 50\,km/s are drawn in blue, odd multiples in red.}
    \label{fig_hrd80}
\end{figure}

Up to a luminosity of $\llso \simeq 6.5$, at about 190\,$M_{\odot}$, 
the effective temperature of the ZAMS increases with increasing luminosity. For higher
initial masses this behaviour changes. The ZAMS moves towards lower effective temperature with increasing
luminosity as a result of stars having significantly increased radii and as a consequence of their proximity to
the Eddington limit. This effect is discussed in more detail in Sect.~\ref{sec3_inf}.

Figure~\ref{fig_hrd} shows only the tracks of the slowest and fastest rotators in our grid.
However, we point out that in the HR-diagram, the evolutionary tracks do not change much
below an initial rotational velocity of $\sim 250\,$km/s. This is demonstrated by the
example of 80$\mso$ models in Fig.~\ref{fig_hrd80}, for which only the tracks 
with initial rotational velocities of $\sim 300\,$km/s and higher deviate significantly from
the track of the model without rotation. As shown in Sect.~\ref{sec3_surf}, this is due to
the absence of significant rotationally induced mixing of helium below the threshold
rotational velocity for quasi-chemically homogeneous evolution (cf. \citet{Brott_2011}).
We note that although the evolutionary tracks of the sequences computed
with initial rotational velocities below $\sim 250\,$km/s are almost identical,
the corresponding models show significant differences concerning the evolution of
the surface abundances of trace elements, e.g. boron and nitrogen (cf. Sect.~\ref{sec3_surf}). 

The models without significant mixing of helium, in the mass range between 60$\mso$ and 80$\mso$,  
expand during core hydrogen burning to surface temperatures as low as 5000\,K. This occurs partly because of
the relatively large amount of convective core overshooting in our models. 
A second reason is that when these models evolve through core hydrogen burning, 
they approach the Eddington limit because their L/M ratio is increasing, and at the same time
their envelope opacity is increasing as cooler surface temperatures are achieved
(cf. Sect.~\ref{sec3_inf}). For higher initial masses, the redward evolution is truncated at
$\Teff \simeq 25\,000\,$K because of the bi-stability mass-loss enhancement in the \citet{Vink_2000} 
mass-loss recipe and the assumed mass-loss enhancement for stars near their Eddington limit
(Sect.~\ref{sec2}).

Figure\,\ref{fig_hrd} shows that the evolutionary tracks of even our slowly rotating models avoid the
upper-right corner of the HR diagram. This is remarkable, since our very massive star models evolve 
very close to their Eddington limit, which leads to an inflation of the envelope (cf. Sect.~\ref{sec3_inf}).
We find that the high mass-loss rates at temperatures below $\sim 30\,000\,$K lead to
significant helium enrichments for all stars above $\sim 60\mso$ (cf. Sect.~\ref{sec3_surf}) such that, as core hydrogen burning continues, they evolve towards hotter rather than cooler surface temperatures.

Figure\,\ref{fig_hrd} also contains the empirical upper luminosity boundary of stars in the Milky Way, as derived
by \citet{Humphreys_1994}. As we see, our slowly rotating models do penetrate the Humphreys-Davidson (HD) limit
and spend a significant amount of time at cooler temperatures. Whether this prediction is in contradiction with
observations for the LMC is currently unclear. The stellar statistics near the HD limit for LMC stars from published
work is not very good \citep{Fitzpatrick_1990}. And even in the Milky Way, stars above the
HD limit are observed, the most prominent example being $\eta\,$Carinae with a luminosity of $\llso = 6.7$ and
$T_{\rm eff}\simeq 30\,000\,$K \citep{Smith_2013}. 

In any case, our models predict a short lived ($\sim 10^5\,$yr) yellow or red supergiant phase of stars in the
mass range $60\mso$ to $80\mso$ during which the core is still burning hydrogen. Our models obtain a stellar wind mass-loss
rate of the order of $10^{-4}\msoy$ during this stage. However, in this part of the HR diagram, mass-loss
rates are very uncertain. We note that in particular higher mass-loss rates (which may be due to the pulsational
instability of these models; Sanyal et al., in prep.) would lead to shorter life times in this evolutionary stage.

The higher the initial mass of our model, the higher is their convective core mass fraction.
While it exceeds 80\% in zero-age main sequence stars above $100\mso$, it reaches 90\%
at $300\mso$, and $500\mso$ stars with a convective core mass fraction of 95\% are almost fully convective. 
This puts the convective core boundary far out into the stellar envelope, where the
pressure scale height is rather small. Consequently, convective core overshooting does not
have the same importance as at smaller stellar mass and plays only a minor role in 
most of the models presented here.

The effect of rotation on the evolution of massive stars is discussed previously, for example in  
\citet{Maeder_2000,Heger_2000b,Brott_2011,Chieffi_2013}. In our models,
there are two main changes in the stellar evolution tracks in the HR diagram. 
First, the effective gravity is reduced as a result of the centrifugal acceleration, which leads to a decrease in the effective temperature 
and luminosity of a star compared to a non-rotating model. Second, above a threshold rotational velocity
for which the timescale of rotational mixing becomes smaller than the nuclear timescale,
the stars evolve quasi-chemically homogeneously \citep{Yoon_2005,Woosley_2006}, and the corresponding models evolve directly towards
the helium main sequence in the HR-diagram \citep{Brott_2011}. 

Both effects are clearly visible in our models. For stellar evolution tracks with initial 
rotational velocities below 250\,km/s, the first effect mentioned is dominant. 
It can be recognized most easily from the ZAMS position of our stellar models 
(Figs.~\ref{fig_hrd} and \ref{fig_hrd80}), where the models are dimmer and cooler the
faster they rotate. 

The most rapidly rotating stellar models undergo quasi-chemically homogeneous evolution. As displayed in Fig.~\ref{fig_hrd}
for $\varv_\mathrm{ZAMS}=400, 500$\,km/s, 
those with initial masses below $\sim 125\mso$ show a strong increase in luminosity and 
effective temperature. For more massive models, however, homogeneous evolution also leads to 
higher luminosities, but the surface temperature is decreased. Towards core helium exhaustion,
strong Wolf-Rayet mass loss leads these stars to lower luminosities and larger surface temperatures
(cf. Sect.~\ref{sec3_mass}).
We return to the discussion of  quasi-chemically homogeneous evolution in Sect.~\ref{sec3_chem}.
Isochrones in the HR-diagram based on tracks discussed here are presented in Appendix~A.


\subsection{Mass loss and surface rotational velocity}
\label{sec3_mass}

\begin{figure*}[htbp]
    \centering
     \includegraphics[angle=-90,width=18cm]{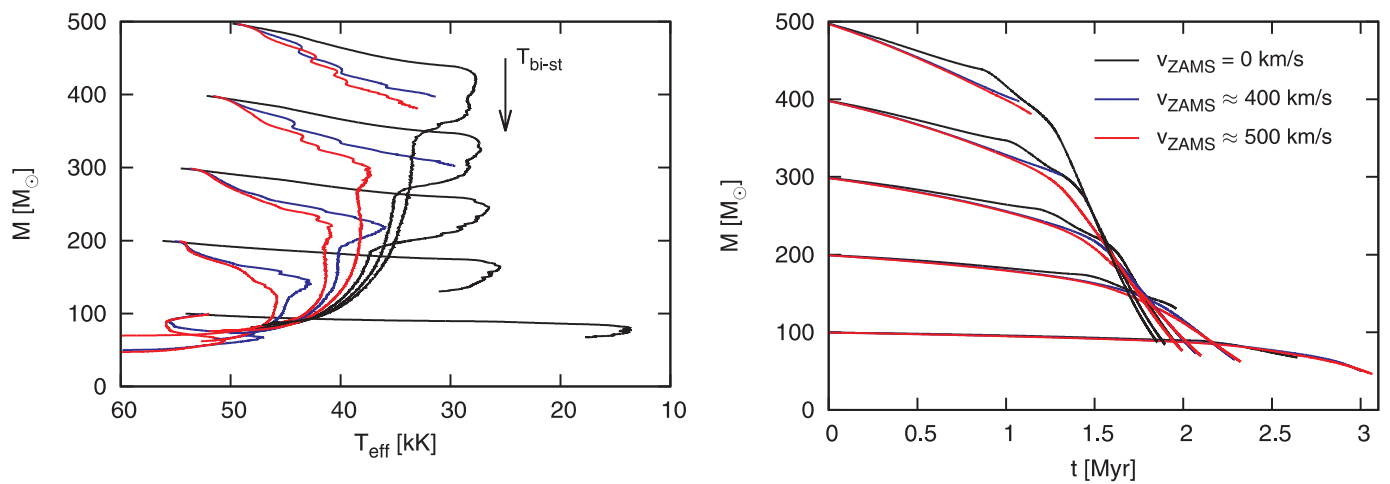}
    \caption{Mass as a function of effective temperature (left panel) and as a function of time (right panel). 
Stellar models that evolve towards effective temperatures below 27\,000\,K are affected 
by the bi-stability jump leading to an increase in the mass-loss rate. This is followed by a 
change in the slope of the mass as a function of time. The increase in mass-loss rate for 
rapidly rotating models is related to the change to the \citet{Hamann_1995} mass-loss recipe.}
    \label{fig_mass}
\end{figure*}

The evolution of the models for the most massive stars is strongly affected by mass loss. 
According to our mass-loss prescription (Sect.~\ref{sec2}), our models undergo mass 
loss of three different strengths. Initially, the mass-loss rate proposed by \citet{Vink_2000, Vink_2001}
is used. For our adopted composition, the mass-loss rate is initially of the order of $3 \cdot 10^{-6}\msoy$
for the $100\mso$ models, while for the $500\mso$ models it is $\sim 6 \cdot 10^{-5}\msoy$.
At this rate, only 10-20\% of the initial mass will be lost over the lifetime
of the stars (Fig.~\ref{fig_mass}, right panel).

The models that do not experience quasi-homogeneous evolution undergo an increase in their mass-loss
rate at effective temperatures of $\sim 25\,000\,$K, which is referred to as the bi-stability
jump \citep{Vink_1999}, according to the Vink et al. prescription (Fig.~\ref{fig_mass}). The corresponding mass loss
is so intense that the models above $100\mso$ stop evolving redward at this stage, since their surfaces 
become strongly helium-enriched. For our most massive models, the 500$\mso$ sequences, the 
maximum mass-loss rate at this stage is about $2 \cdot 10^{-4}\msoy$.

Once the surface helium mass fraction has increased owing to mass loss to more than 40\%,
the Wolf-Rayet mass-loss rate is phased in, reaching its full strength at a surface helium mass fraction
of 70\%. As a consequence, our $100-500\mso$ models lose  50-80\%  of their initial mass
before core hydrogen exhaustion. During the phase of Wolf-Rayet winds, the models evolve at
decreasing luminosity and all accumulate in a narrow region in the HR-diagram (Fig.~\ref{fig_hrd}).

The three different phases of mass loss can be clearly seen in both panels of Fig.~\ref{fig_mass}.
Again, for our most extreme models, the slowly rotating stars with an initial mass of $\sim 500\mso$,
the largest obtained mass-loss rate is $\sim 5 \cdot 10^{-4}\msoy$, according to our Wolf-Rayet mass-loss prescription.
At this moment, the stars have a luminosity of $\llso \simeq 7.1$.
Assuming a terminal wind speed of $1000\,$km/s, we compute a wind-momentum-to-photon-momentum ratio of $\eta \simeq 2$ for this situation. With the same numbers, we obtain
a wind-kinetic-energy-to-luminosity ratio of $0.003$. The corresponding wind darkening is therefore 
negligible \citep{Heger_1996} and significantly smaller than in 
Galactic Wolf-Rayet stars \citep{Lucy_1993}.

The models that evolve quasi-homogeneously first increase their luminosity more than the models described above.
Consequently, their mass-loss rate becomes larger than that of the inhomogeneous models. The homogeneous models
do not reach the bi-stability limit as the increased helium surface abundance due to rotational mixing
keeps their surface temperature above $25\,000\,$K. However, they can reach the Wolf-Rayet stage earlier,
and in the end lose similar amounts of mass as the inhomogeneous models. During the Wolf-Rayet stage,
there are no clear characteristics that can distinguish between the two types of evolution. 

\begin{figure}[htbp]
    \centering
     \includegraphics[angle=-90,width=8.5cm]{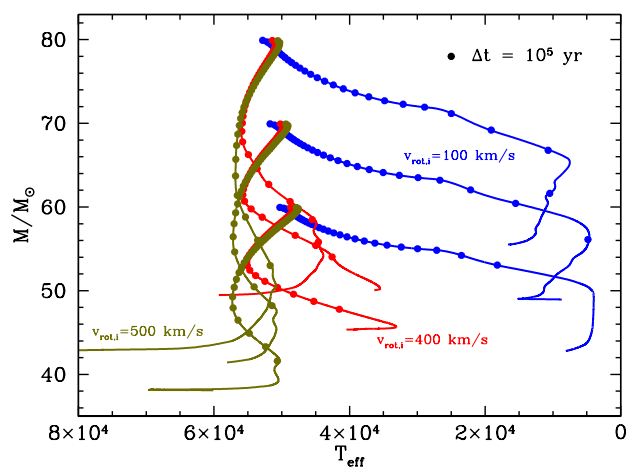}
    \caption{Mass as a function of effective temperature for the models with initial masses
of 60$\mso$, 70$\mso$, and 80$\mso$, and for initial rotational velocities of 100, 400,
and 500\,km/s, as indicated. The time difference between two dots along each of the tracks
is $10^5\,$yr.
}
    \label{fig_m-t}
\end{figure}

Figure~\ref{fig_m-t} compares the mass evolution of the homogeneously and inhomogeneously evolving
60$\mso$ to 80$\mso$ models. It shows that in this mass range, the homogeneously evolving models
lose more mass, with more than half of the mass being lost during the Wolf-Rayet phase.
The inhomogeneously evolving models lose almost equal amounts during their evolution through
the hot part of the HR-diagram as at cool surface temperatures, where the latter mass loss
occurs on a timescale of only $10^5\,$yr. 

\begin{figure}[htbp]
    \centering
     \includegraphics[angle=-90,width=8.5cm]{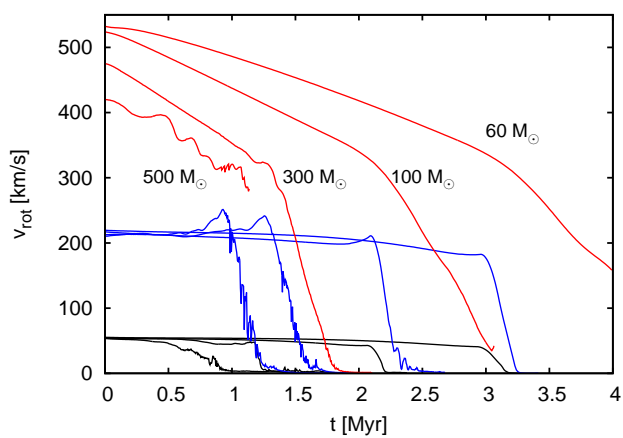}
    \caption{The surface rotational velocity is shown as a function of time for stellar model sequences 
with four different initial masses (as indicated), and for three initial surface rotational velocities.
The rapid decrease in rotation rate seen for $\varv_\mathrm{i} \simeq 50, 200$\,km/s is 
caused by the enhanced mass loss at the bi-stability jump, whereas the steep decline of the rotation velocity
for the fast rotators ($\varv_\mathrm{i}\simeq 500$\,km/s) is a consequence of Wolf-Rayet type mass loss. 
}
    \label{fig_velo}
\end{figure}

As a consequence of the initially rather moderate stellar wind mass loss, most of our models
evolve at a constant rotational velocity for most of the core hydrogen burning. 
In this case the loss of angular momentum through the stellar wind \citep{Langer_1998}
and the increased momentum of inertia due to expansion is compensated by the transport
of angular momentum from the contracting core to the envelope. As shown in Fig.~\ref{fig_velo},
this holds even for the most massive stars in our grid. Only the models which undergo chemically
homogeneous evolution, as a result of their increased luminosity and mass-loss rate, show a moderate decline
of their surface rotational velocity before entering the phase of Wolf-Rayet winds. 

The strong mass loss for $T_{\rm eff} < 25\,000\,$K for the inhomogeneously evolving models, and the
Wolf-Rayet type mass loss for the homogeneously evolving ones both lead to a strong spin-down of the
stars towards the end of their core-hydrogen burning evolution. In fact, this is only avoided
for stars initially below $\sim 30 \mso$ (cf. Fig.~3 of \citet{Vink_2010}).
Since the magnetic coupling in our models ensures close-to-rigid rotation during core hydrogen
burning, this implies that all our LMC models above $\sim 80\mso$ lose so much angular momentum that
they cannot be considered to produce candidates for long-duration gamma-ray bursts. In that respect,
models that undergo chemically homogeneous evolution of lower mass are better suited (cf. Sect.~\ref{sec3_chem}).  

Effective temperatures and radii of observed massive stars may be affected 
by optically thick stellar winds, making the star appear larger and cooler.
We use Eq.~(14) from \citet{Langer_1989} 
\begin{equation}
\tau(R)=\frac{\kappa |\dot{M}|}{4 \pi R \left( \varv_{\infty} - \varv_0 \right)} \ln{\frac{\varv_{\infty}}{\varv_0}}
\label{eq_tau}
\end{equation}
to estimate the optical depth of the stellar winds of our stellar models. 
Here, $R$ designates the radius of the stellar model without taking the wind into account.
This equation is derived from a $\beta$-velocity law with $\beta=1$.
In this case, we use the electron scattering opacity
($\kappa= \sigma \left( 1+X \right)$, where $\sigma$ is the Thomson scattering cross section), 
an expansion velocity $\varv_0=20$\,km/s at the 
surface of the stellar model, and a terminal wind velocity of $\varv_{\infty}=2000$\,km/s.

\begin{figure}[htbp]
    \centering
     \includegraphics[angle=-90,width=8.5cm]{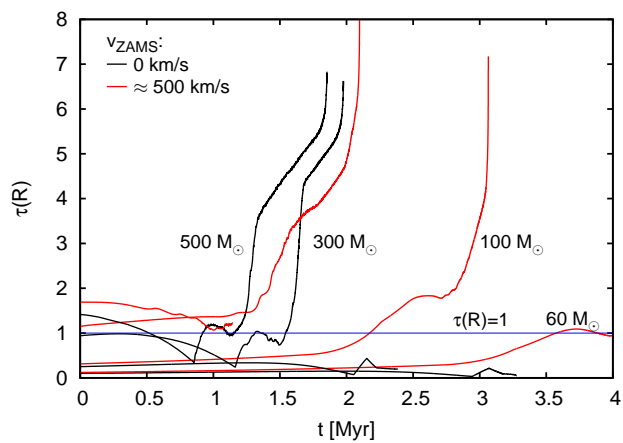}
    \caption{Stellar wind optical depth according to Eq.~(1), for some of our model sequences, 
shown as a function of time. 
We show eight stellar models with four different initial masses (60, 100, 300, 500\,$M_{\odot}$), and two initial surface 
rotational velocities (0, 500 km/s) in black and red, respectively. The line for unit wind optical depth is plotted to facilitate the comparison.}
    \label{fig_thick}
\end{figure}

Figure~\ref{fig_thick} shows the estimated optical depth of stellar winds as a function of time for several model sequences. The 
behaviour of the optical depths seen in the figure is mostly related to the change 
in the mass-loss rate. The optical depth increases for higher initial mass and higher
rotation rate as a result of a corresponding increase in the mass-loss rate. 
While these numbers are only approximations, it shows that the winds of the most massive
stars might already be optically thick ($\tau > 1$) on the zero-age main sequence,
which implies that spectroscopically, these stars may already show Wolf-Rayet characteristics at this time (see also \citet{Graefener_2008,Crowther_2010}). 
Furthermore, while the stars of $100\mso$ and below are expected to show optically thin winds
for most of the core hydrogen burning evolution, the optical depths of the winds for models for which 
a Wolf-Rayet type wind has been assumed may be quite large. 

\subsection{Surface abundance evolution}
\label{sec3_surf}
Deriving masses or ages of stars using stellar evolution tracks or isochrones in the HR diagram is no longer straightforward when 
rotational mixing is important. A given pair of effective temperature and luminosity can be explained by 
several combinations of masses, rotational velocities, and ages. To uniquely determine the initial parameters and the age 
of a star, it is necessary to include additional observables. The mass fractions or abundances of elements at the stellar surface 
are tracers of the rotational mixing. They can be used to uniquely identify the initial parameters and the age 
of a star in a three (or more) dimensional space of observables (cf. \citet{Koehler_2012}). 
We therefore discuss the influence of time, initial rotational velocity and 
initial mass on the surface composition of our models.

\begin{figure*}[htbp]
    \centering
     \includegraphics[angle=-90,width=18cm]{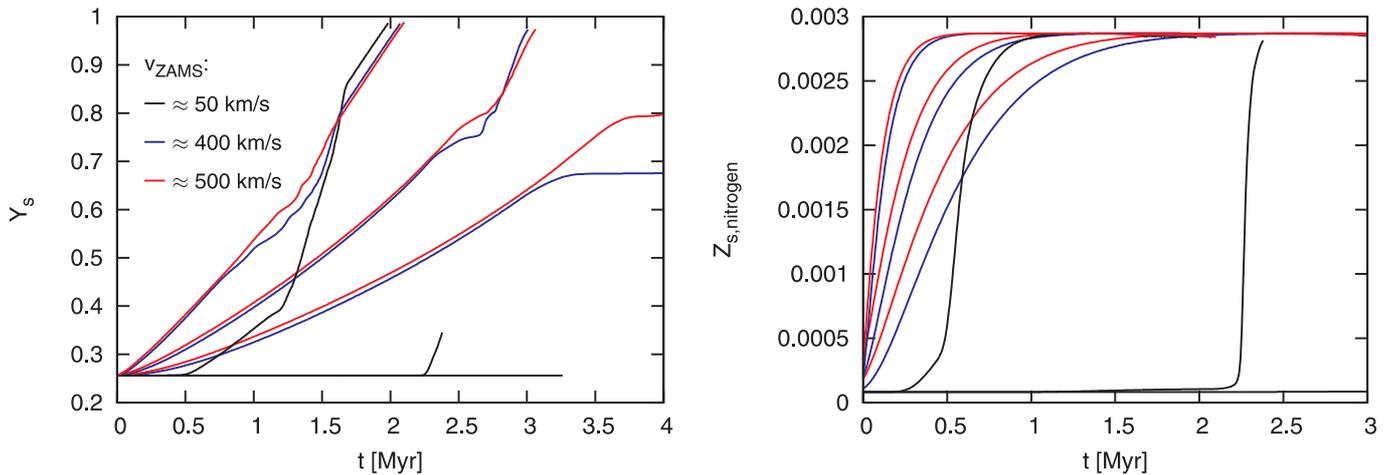}
    \caption{Helium (left panel) and nitrogen (right panel) surface mass fraction as a function of time 
for models of 60, 100, and 300\,$M_{\odot}$ and rotation rates explained in the figure key. 
While for the fast rotators the enhancements are mostly due to rotational mixing, 
for the slowly rotating models the increase in helium and nitrogen in the 100\,$M_{\odot}$ and 300\,$M_{\odot}$
mass models is solely due to mass loss.}
    \label{fig_NH}
\end{figure*}   

\begin{figure*}[htbp]
    \centering
     \includegraphics[angle=-90,width=8cm]{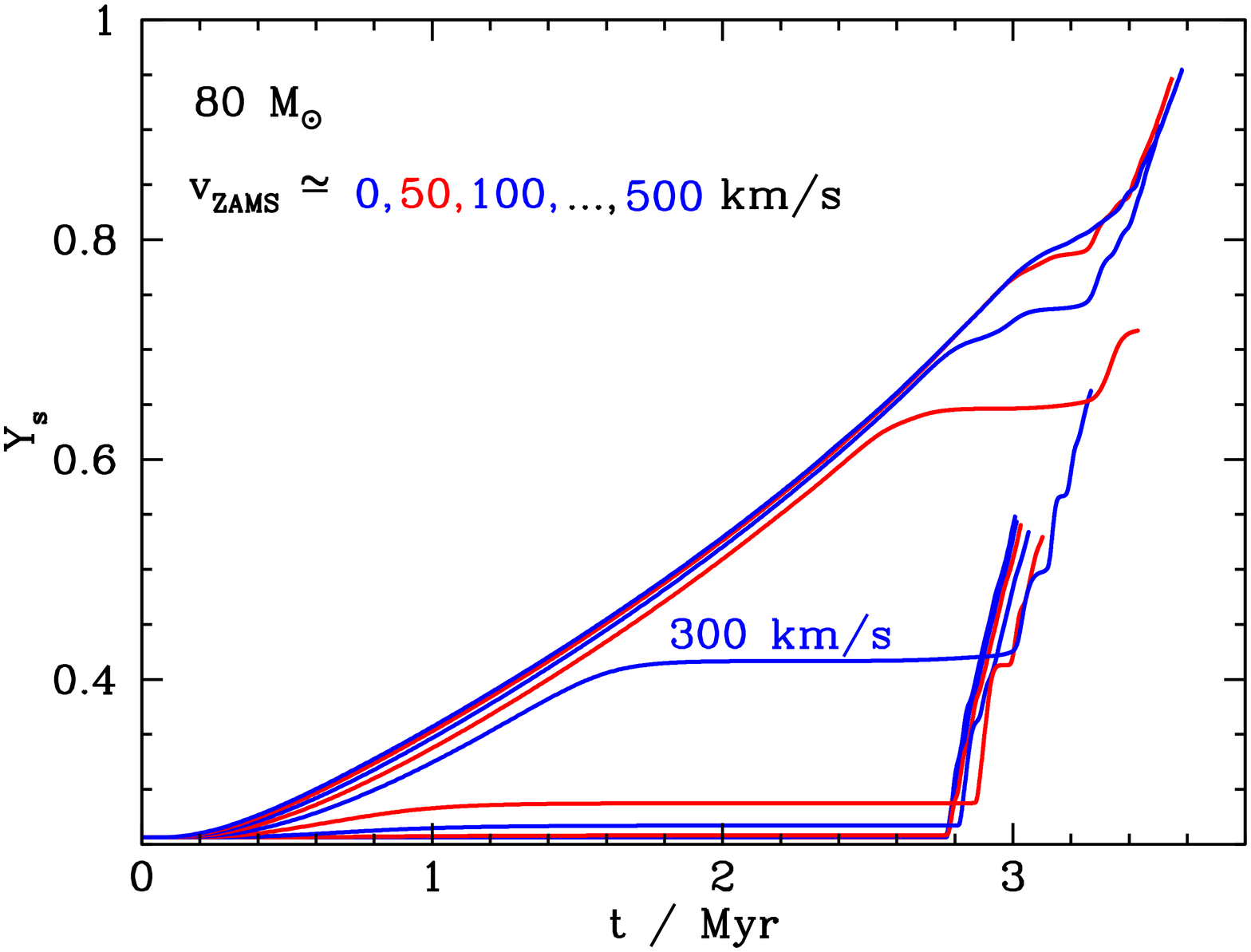}
     \includegraphics[angle=-90,width=8cm]{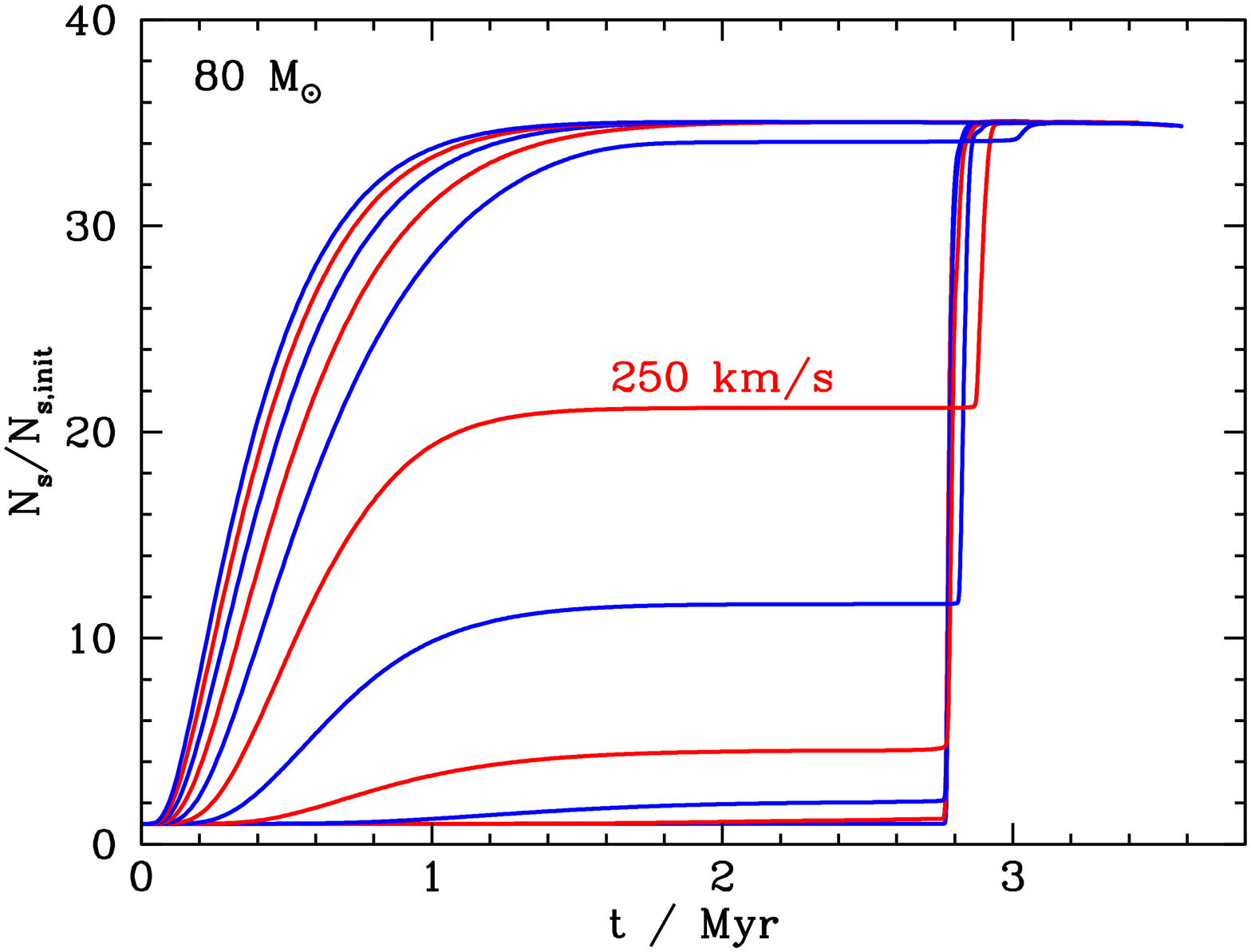}
    \caption{Helium (left panel) and nitrogen (right panel) surface mass fraction as a function of time for models
of 80 \,$M_{\odot}$ with approximate initial rotational velocities of
0, 50, 100, 150, 200, 250, 300, 350, 400, 450, and 500~km/s, during their core hydrogen burning evolution.}
    \label{fig_NH1}
\end{figure*}

\begin{figure*}[htbp]
    \centering
     \includegraphics[angle=-90,width=12.23cm]{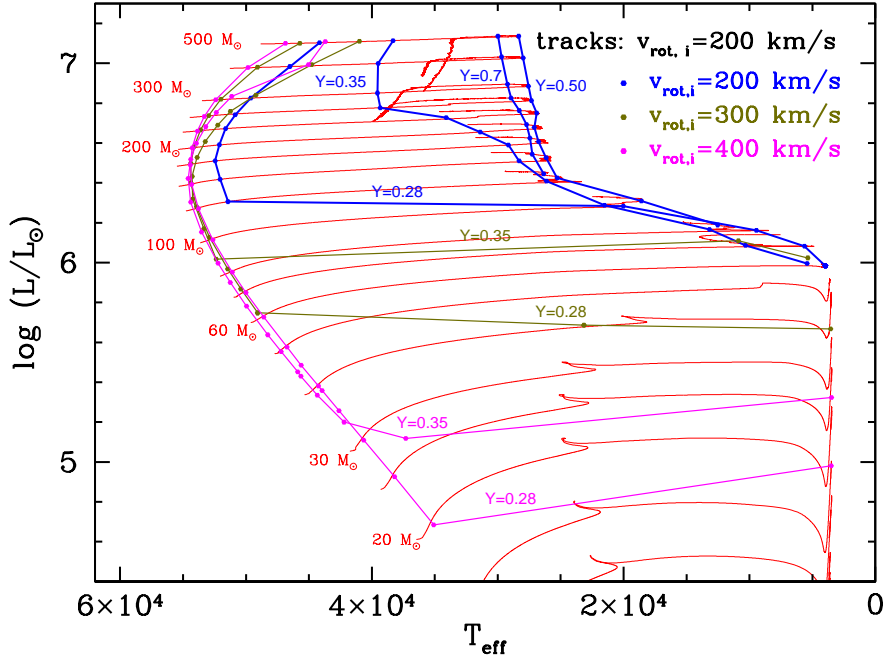}
    \caption{Evolutionary tracks in the HR diagram of stellar models initially rotating with approximately 200~km/s,
     and with initial masses of 12, 15, 20, 25, 30, 40, 50, 60, 70, 80, 100, 125, 150, 175, 200, 230, 260, 
     300, 400, and 500$\mso$. Overlaid are lines (in blue) of constant helium surface mass fraction for Y=0.28, 0.35, 0.50, and 0.70. Lines of constant helium surface mass fraction for Y=0.28, 0.35 corresponding to
     models with approximate initial rotational velocities of 300 and 400~km/s are also shown.}
    \label{fig_NY}
\end{figure*}

\begin{figure*}[htbp]
    \centering
     \includegraphics[angle=-90,width=12.23cm]{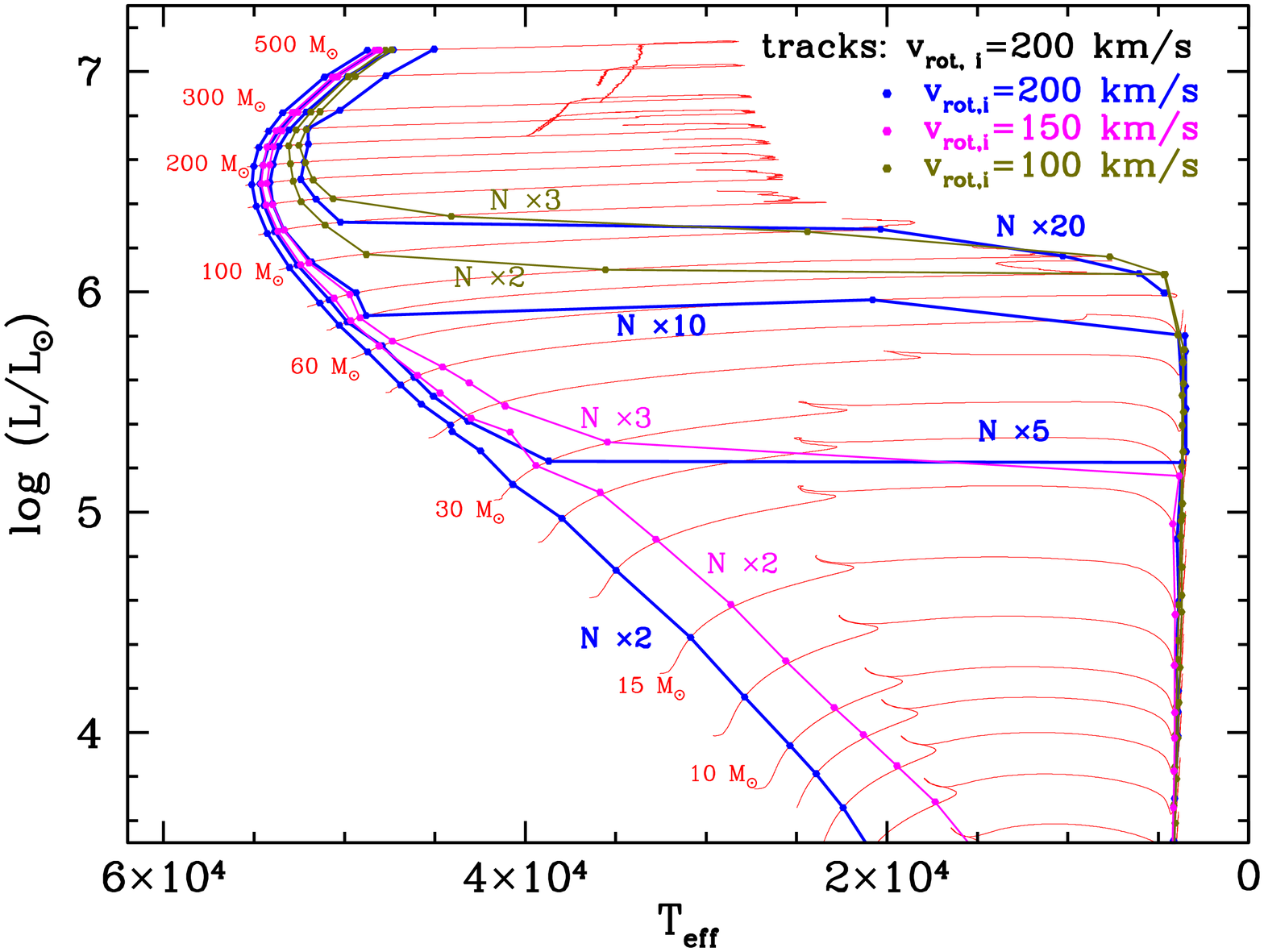}
    \caption{Evolutionary tracks in the HR diagram of stellar models initially rotating with approximately 200~km/s,
     and with initial masses of 6, 7, 8, 9, 10, 12, 15, 20, 25, 30, 40, 50, 60, 70, 80, 100, 125, 150, 175, 200, 230, 260,
     300, 400, and 500$\mso$. Overlaid are lines (in blue) of constant nitrogen surface mass fraction, corresponding to nitrogen
     enhancement factors of 2, 5, 10, and 20, 
     as indicated in blue. Lines of constant nitrogen
     enhancement factors of 2 and 3 corresponding to 
     models with approximate initial rotational velocities of 150 and 100~km/s are also shown.}
    \label{fig_NH2}
\end{figure*}

The rotational mixing in our models is sensitive to the gradient of the mean molecular weight $\mu$.
Essentially, any significant $\mu$-barrier prevents rotational mixing. Therefore, mixing of helium
can only occur in our models as long as they are quasi-chemically homogeneous. This, in turn, requires
the timescale for rotational mixing to be shorter than the nuclear timescale, as the star attempts to establish a 
$\mu$-barrier on the nuclear timescale.

In the fastest rotators, e.g. in the $100\mso$ and $300\mso$ models initially rotating with
$\sim 400$ and 500\,km/s, the surface helium abundance goes almost all the way to $Y_{\rm s}=1$
towards core hydrogen exhaustion (cf. Fig.~\ref{fig_NH}). However, the same figure shows that this
evolution is truncated for the rapidly rotating 60$\mso$ models, at $Y_{\rm s}=0.68$ and $Y_{\rm s}=0.80$ 
for initial rotational velocities of $\sim 400$ and 500\,km/s, respectively. These models spin down 
during core hydrogen burning (Fig.~\ref{fig_velo}) such that the timescale for rotational mixing
eventually becomes larger than the nuclear timescale. 

Figure~\ref{fig_NH} shows that the non-rotating $300\mso$ sequence also evolves to $Y_{\rm s}=1$.
The reason is the large convective core fraction of this model, and its large mass-loss rate.
Whereas the model always keeps a small radiative envelope mass, the equivalent amount of mass 
is lost on a short timescale such that core and surface abundances become almost equal
(cf. Sect.~\ref{sec3_chem}; Eq.\,2).
 For the rapidly rotating $100\mso$ and $300\mso$ models, mass loss rather than rotational mixing
must take over to enforce chemical homogeneity, since these models also spin down quite dramatically
(Fig.~\ref{fig_velo}). 

Figure~\ref{fig_NH1} (left panel) demonstrates the role of rotational mixing of helium in greater detail 
using the example of 80$\mso$ sequences. In the models initially rotating with 0, 50, 100, 150, and 200\,km/s,
such mixing is essentially negligible, and only mass loss can increase the helium surface abundance
close to core hydrogen exhaustion, which is why all of these models evolve along the same
track in the HR-diagram (Fig.~\ref{fig_hrd80}). The models between 250 and 400\,km/s start out
with chemically homogeneous evolution (though the one with $\varv_{\rm rot,i}=250\,$km/s only for a short 
amount of time), and truncate the homogeneous evolution later in time for higher initial rotation.
The models which initially rotate with about 450 and 500\,km/s undergo chemically homogeneous evolution
all the way (though with the help of mass loss in the end), and their evolution in the HR diagram
is also practically identical (Fig.~\ref{fig_hrd80}).

The evolution of the surface nitrogen abundance in our models of rotating and mass losing stars follows
different rules. During a phase of chemically homogeneous evolution, nitrogen is quickly mixed from
the core to the surface establishing the CNO-equilibrium value in atmospheric layers (Figs.~\ref{fig_NH} 
and~\ref{fig_NH1}, left panels). Mass loss can also lead to nitrogen enhancements, even without
rotational mixing (Fig.~\ref{fig_NH}). Figure~\ref{fig_NH1} shows that a strong nitrogen surface enhancement
can be obtained even in models with initial rotational velocities well below the threshold value required
for chemically homogeneous evolution. For example, the 80$\mso$ models with $\varv_{\rm rot,i}=150\,$ and $200\,$km/s 
enrich nitrogen at the stellar surface by factors of~4 and~12 by rotation, i.e. before mass loss kicks in. This order of magnitude of nitrogen enrichment is as expected, since our models were calibrated to increases the surface nitrogen abundance by a factor of about 3 for stars of 13--15$\mso$ for an initial rotational velocity of 150\,km/s, and rotational mixing is stronger in more massive stars.

The reason is that nitrogen, carbon, and oxygen, with a maximum mass fraction of less than one per cent, are just trace elements, 
such that even significant internal gradients do not lead to strong $\mu$-barriers.
Therefore, at the beginning of core hydrogen burning, before significant amounts of hydrogen
have been converted to helium, but after the CNO-cycle has already strongly enhanced the nitrogen abundance in the
convective core, rotational mixing can bring nitrogen out of the core into the radiative envelope
and later on also to the surface. 

The surface enrichment of helium and nitrogen explained above allows a simple understanding
of the occurrence of surface enrichments in the HR diagram. As shown in Fig.~\ref{fig_NY},
stars initially rotating with rotational velocities of 200\,km/s or less that are less luminous than
$\llso\simeq 6.2$ are not expected to show
any helium surface enrichment (i.e. $Y_{\rm s} < 0.28$) as long as their surface temperature is higher than $\sim 20\,000\,$K.
The same is true for stars initially rotating slower than 300\,km/s below $\llso\simeq 5.6$, and for those
initially rotating slower than 400\,km/s below $\llso\simeq 4.8$. 
Similar higher thresholds can be read off from Fig.~\ref{fig_NY} for larger surface helium mass fractions.
Given that \citet{Agudelo_2013} found that 75\% of all O~stars rotate slower than 200\,km/s
and that amongst the 31 O2 to O5 stars none was found to rotate faster than $\sim 300\,$km/s,
the helium enrichment in LMC stars below $\sim 100\mso$ is expected to be quite small
during the main-sequence phase.

In contrast, as more than half of all O stars were found to rotate faster than $100\,$km/s, 
nitrogen enrichment by at least a factor of 2 is expected to be almost ubiquitous above
$\llso\simeq 6.0$, and quite frequent in main-sequence O stars in general (Fig.~\ref{fig_NH2}).

Changes in the amount of neon, sodium, aluminum, and magnesium at the surface also occur,
indicating that in very massive stars the MgAl- and the NeNa-cycles are active. 
In our models, the amount of magnesium is reduced while aluminum 
is produced. The mass fraction of neon increases in our models, while 
sodium decreases. For more details, we refer to the electronic data published with this paper. 

Lithium, beryllium, and boron are destroyed in the stellar interior \citep{McWilliam_2004} 
at temperatures above $\sim 2.5 \cdot 10^{6}$\,K for lithium, $\sim 3.5 \cdot 10^{6}$\,K for beryllium, and 
$\sim 5.0 \cdot 10^{6}$\,K for boron. In our models, temperatures below $5\cdot 10^{6}$\,K are 
only found in the outer envelope. 
Therefore lithium, beryllium, and boron can exist in the outer envelope, while they are destroyed in deeper layers. 
For the slowest rotating models, the surface abundances of these three elements remain constant 
until layers which were exposed to higher temperatures earlier in the evolution are exposed to the surface due to mass loss. 
When that happens, the mass fractions of beryllium, boron, and lithium are quickly reduced. 
For fast rotating models, rotational mixing and mass loss 
lead to a gradual decrease in lithium, beryllium, and boron over time.

\subsection{Chemically homogeneous evolution}
\label{sec3_chem}

\begin{figure}[htbp]
    \centering
     \includegraphics[width=8.5cm]{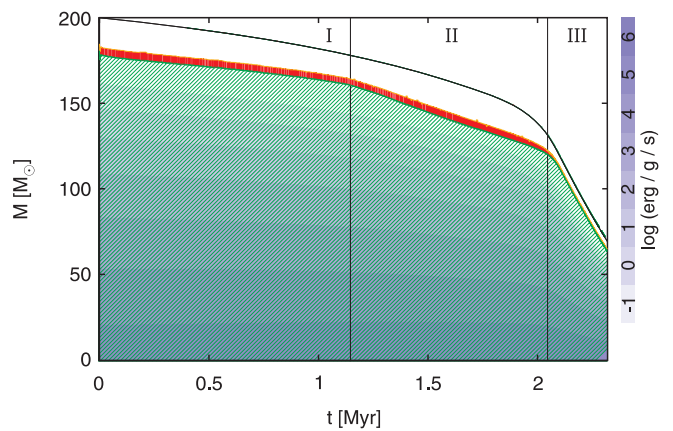}
    \caption{Kippenhahn-diagram for our 200\,$M_{\odot}$ model with $v_\mathrm{ZAMS}\simeq 300$\,km/s. The black solid line gives the stellar mass
as a function of time. Blue shading indicates thermonuclear energy generation (see colour bar to the right 
of the plot). Green hatched parts show convective regions, and convective core overshooting is indicated in red. 
Three different regimes can be distinguished according to the rate at which the mass of the convective core decreases (see text).}
    \label{fig_kippenhahn}
\end{figure} 

\begin{figure}[htbp]
    \centering
     \includegraphics[angle=-90,width=8.5cm]{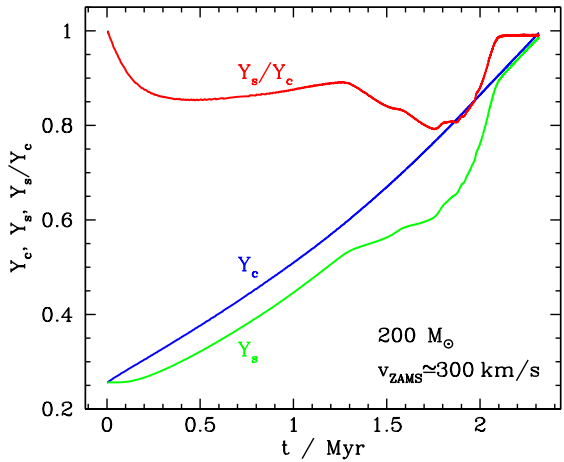}
    \caption{Central $Y_{\rm c}$ and surface $Y_{\rm s}$ helium abundance of our 200\,$M_{\odot}$ model with $v_\mathrm{ZAMS}\simeq 300$\,km/s
(cf. Fig.~\ref{fig_kippenhahn}), and the ratio $Y_{\rm s}/Y_{\rm c}$ as a function of time.}
    \label{fig_ycys}
\end{figure}

As mentioned in Sect.~\ref{sec3_trac}, we can divide our models into three classes.
The first one (which we call Class~O) likely corresponds to most stars in observed stellar samples and contains
the models that evolve in the normal way, i.e. in which the rotationally induced mixing of helium
is negligible. The second one describes the models that undergo quasi-chemically homogeneous evolution (Class H), which correspond
to the initially fastest rotators. As the third class, we have the models that start out evolving homogeneously,
but which spin down such that the rotational mixing of helium stops (Class HO), after which they evolve
redward in the HR diagram as the ordinary models. These models retain a memory of their past homogeneous evolution,
in that they will keep an enhanced helium surface abundance and a higher luminosity-to-mass ratio \citep{Langer_1992}
compared to ordinary stars throughout the rest of their core hydrogen burning evolution.

At the highest masses considered here ($M \simgr 150\mso$), our models may also undergo quasi-chemically homogeneous evolution without
rotationally induced mixing, which is due to a combination of an extremely high fraction of the convective core to the total 
stellar mass and a very high Wolf-Rayet type mass-loss rate. We illustrate this by the example of 
our 200\,$M_{\odot}$ sequence with $v_\mathrm{ZAMS}\simeq 300$\,km/s in Fig.~\ref{fig_kippenhahn}.
For such very massive stars, the convective core comprises a major fraction of the stellar mass.

We can divide the core hydrogen burning evolution of this model into three parts according to the mass evolution of the convective core.
During the first part, the mass of the convective core decreases as a function of time, but more slowly than the mass of the star,
such that the mass of the radiative envelope decreases and the convective core mass fraction increases. 
During this phase, the model undergoes quasi-chemically homogeneous evolution (cf. Fig.~\ref{fig_ycys}). 
At $t\simeq 1.2\,$Myr, the model transitions to ordinary evolution, from which time on the convective core mass
decreases somewhat faster than the total mass. During this second part of its evolution, the helium surface
abundance increases, even though at a much lower rate than the central helium abundance (Fig.~\ref{fig_ycys}).
Finally, in part three, the Wolf-Rayet mass loss kicks in, which leads to a very small mass of the non-convective
stellar envelope ($M_{\rm env}\simeq 5\mso$). The amount of mass corresponding to the non-convective envelope is
lost (at a rate of $\sim 2\, 10^{-4}\,\msoy$) on a timescale of $\sim$20\,000\,yr
which corresponds to only about 1\% of the core hydrogen burning time, $\tau_{\rm H}$. Consequently, the surface helium mass fraction
during this stage is roughly  equal to (within 1\%) the central helium mass fraction. Thus, regardless of rotation,
the model is extremely chemically homogeneous during this phase.   

We define a fourth type of evolution (Type~M) by
\begin{equation} 
{M_{\rm env}\over {\dot M}} < 0.1 \tau_{\rm H},
\end{equation}
where $\tau_{\rm H}$ is the hydrogen burning timescale.
Since this condition ensures that the surface-to-central helium abundance ratio 
is above $\sim 0.9$ --- i.e. similar to or smaller than in the
case of rotationally induced quasi-chemically homogeneous evolution --- we can subdivide our models into further classes. For example, the 200\,$M_{\odot}$ sequence discussed here can be considered of Class~HOM, as it  first undergoes chemically homogeneous evolution 
due to rotational mixing, then ordinarily, and finally chemically homogeneous evolution due to mass loss.
 
\begin{figure}[htbp]
    \centering
     \includegraphics[angle=-90,width=8.5cm]{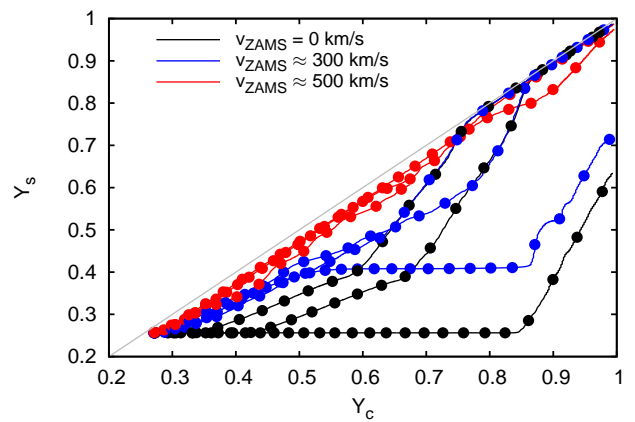}
    \caption{The mass fraction of helium in the stellar core $Y_\mathrm{c}$ and at the surface $Y_\mathrm{s}$ are 
depicted for stellar models of 100, 300, and 500\,$M_{\odot}$ 
and rotation rates explained in the figure key. For a given rotation rate, initially more massive stars
increase their surface helium abundance more quickly.
Every 10$^5$\,yr the models are highlighted by filled circles. Chemically homogeneous evolution is indicated by equal changes in both mass fractions and corresponds to a slope unity. 
At $Y_\mathrm{c} \ge 0.8$, all calculated stellar evolution models show an evolution toward the line of 
equal $Y_\mathrm{c}$ and $Y_\mathrm{s}$, caused by mass loss.}  
    \label{fig_HCS}
\end{figure}

It is instructive to consider Fig.~\ref{fig_HCS} to understand which evolutionary classes are realized by our models.
The non-rotating sequences with initial masses of 300$\mso$ and 500$\mso$ achieve chemical homogeneity at a central
helium mass fraction of 0.85 and 0.75, respectively, and thus belong to Class~OM. The non-rotating 100$\mso$ sequence
does not reach homogeneity. Consequently, although it achieves a quite high final surface helium mass fraction of $\sim$0.65,
it belongs to Class~O. The most rapidly rotating models depicted in Fig.~\ref{fig_HCS} never show a significant discrepancy between
their central and surface helium abundances, which implies that they evolve from H-type to M-type evolution and
belong to Class~HM. The stars rotating initially with $\sim 300\,$km/s all undergo an H$\rightarrow$O transition, where the two
more massive ones move further on to Type~M evolution. As stars in the Classes~HOM and HM both start out and end their chemically homogeneous evolution, we will not distinguish them further and designate them both as Class~HM. 

\begin{figure*}[htbp]
    \centering
     \includegraphics[angle=-90,width=18cm]{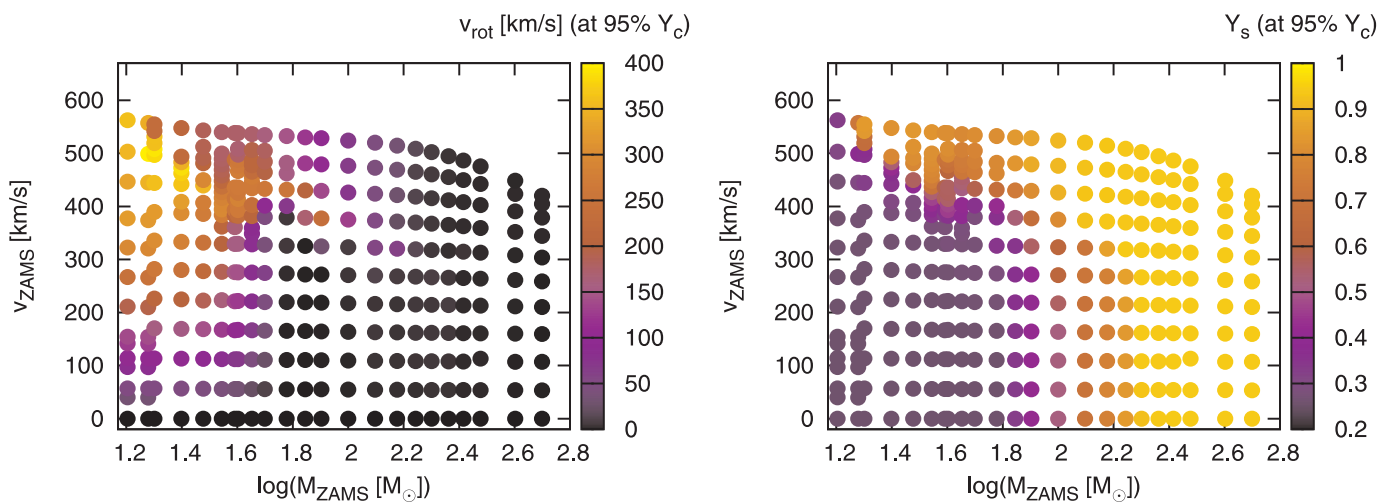}
    \caption{The grid of all initial masses $M_{\mathrm{ZAMS}}$ and surface rotational velocities 
$v_{\mathrm{ZAMS}}$ is shown, including the stellar models with initial masses above 19\,$M_{\odot}$ published 
in \citet{Brott_2011}. The colour coding corresponds to the surface rotational velocity (left panel) and the 
helium mass fraction at the surface (right panel) at the time when the central  helium mass fraction has reached 95\%.}
    \label{fig_homog}
\end{figure*} 

Thus, we have five different classes of evolution, H, HM, HO, OM, and O, where all but the last involve
quasi-chemically homogeneous evolution. Class~O can be identified in Fig.~\ref{fig_homog} 
(right panel), which shows
the close-to-final surface helium mass fraction for all our model sequences. Looking at the slowly rotating
models, this figure shows that mass loss starts affecting the surface helium abundance above an initial mass of
$\sim 65\mso$ ($\log M/\mso \simeq 1.8$), and that M-type chemically homogeneous evolution is obtained above
$\sim 160\mso$ ($\log M/\mso \simeq 2.2$). According to our definition, the slow rotators below $\sim 160\mso$
are thus in Class~O, the more massive ones in Class~OM. For the fast rotators, the dividing line between
homogeneous and inhomogeneous evolution depends on stellar mass. At 20$\mso$ ($\log M/\mso \simeq 1.3$),
stars rotating initially faster than $\sim 500\,$km/s evolve homogeneously (Class~H), while at 
$\sim 160\mso$ the critical velocity is at $\sim 350\,$km/s, and more massive stars above this velocity
are in Class~HM. Below the dividing line defining Class~H, between $20\mso$ and $160\mso$ is a stretch
of Class~HO models, which comprises a larger initial mass range for higher initial mass.

As mentioned in Sect.~\ref{sec3_trac}, rapidly rotating stellar models with initial masses greater than 125\,$M_{\odot}$ show significantly different behaviour in the HR diagram (see Fig.~\ref{fig_hrd}) 
to the stellar models with $M \le 125\,M_{\odot}$ (see also Sect.~\ref{sec3_inf}). Nevertheless, they behave in a similar way in the 
Kippenhahn diagram and when comparing the helium mass fraction in the core and at the surface. 

Stellar evolution models above 80\,$M_{\odot}$ have strong mass loss. Even the fastest rotators are slowed 
down significantly when the \citet{Hamann_1995} mass-loss rate is applied. For models above 
$M_{\mathrm{ZAMS}}\simeq 100\,M_{\odot}$, the rotation rate at 95\% helium mass fraction in the core is below 
100\,km/s. The surface rotational velocity goes down to less than 50\,km/s for models above 150\,$M_{\odot}$, 
independent of the initial surface rotational velocity (Fig.~\ref{fig_homog}; right panel). 

From the above, we can come to several conclusions which are relevant in comparison to observed stars.
First, helium-enriched single stars below $\sim 65\mso$ are prime suspects of H- or HO-type chemically homogeneous
evolution, whereas M-type evolution can be excluded for them. These stars are expected to preserve
their rapid rotation throughout their core hydrogen burning evolution.
As a consequence, we may expect some correlation of the helium surface abundance with the stellar
rotation rate in the considered mass regime. 

For very massive stars (above $\sim 200\mso$) rotation, whether initially fast or not, makes little difference,  i.e. the evolution becomes almost independent of the initial rotational velocity. These stars are
efficiently spun down by mass loss, and all undergo chemically homogeneous evolution during their
advanced hydrogen-burning evolution.  Their surface helium abundance is always very close
to their central helium abundance because of a rapid loss of the thin outer radiative envelope through
mass loss.
As a consequence, these sequences can produce models that have a very
high surface helium mass fraction ($Y_{\rm s} > 0.9$) and are still undergoing  
core-hydrogen burning.

\subsection{Mass-luminosity relation}
\label{sec3_ml}

\begin{figure}[htbp]
    \centering
     \includegraphics[angle=-90,width=8.5cm]{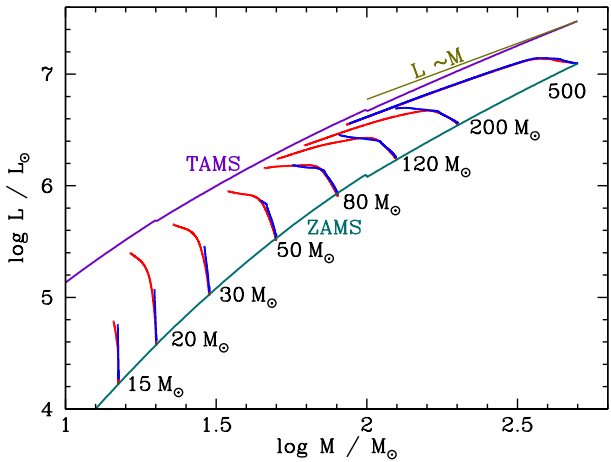}
    \caption{Luminosity as a function of mass for selected non-rotating (blue) and rapidly rotating models
($v_{\rm rot,i}=500\,$km/s; red).
The labels indicate the initial mass of the considered sequences. The tracks end at a central helium mass fraction of
$Y_c=0.98$.
Overplotted are the mass-luminosity relations of \citet{Graefener_2011}
for chemically homogeneous stars with a hydrogen mass fraction of X=0.74 (labelled ZAMS) and X=0 (labelled TAMS).
The straight line labelled ``$L\sim M$'' indicates the smallest expected slope of the mass-luminosity relation.  
}
    \label{fig_ml}
\end{figure}

Stars of higher mass are increasingly luminous. The most massive models in our grids, at $500\mso$,
radiate at more than $10^7\lso$. 
Figure~\ref{fig_ml} shows the evolution of selected non-rotating and of rapidly rotating model sequences
in the mass-luminosity plane, from core hydrogen ignition to core hydrogen exhaustion.
At the lowest considered masses ($15 \mso$), the models evolve vertically upward since
their mass loss is negligible. The more massive, rapidly rotating models evolve to higher luminosity, and they
turn left toward lower mass.

A comparison with the mass-luminosity relations for chemically homogeneous stars of \citet{Graefener_2011}
reveals that, except for the depicted $15 \mso$ model, the rapidly rotating sequences
shown in Fig.~\ref{fig_ml} undergo quasi-chemically homogeneous evolution. They start at the mass-luminosity (ML) relation
for zero-age main sequence stars and finish close to the ML-relation for chemically homogeneous helium stars,
indicating their nearly homogeneous chemical structure.
This also demonstrates that the agreement of the ML-relations of \citet{Graefener_2011} (which represent
fits to stellar models computed with a different code) with our models is very good.

\begin{figure}[htbp]
    \centering
     \includegraphics[angle=-90,width=8.5cm]{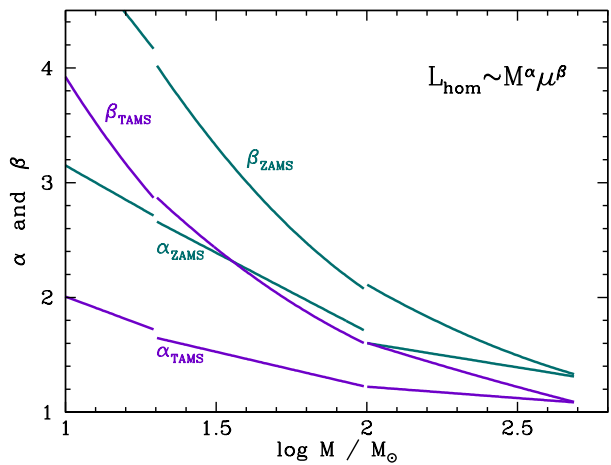}
    \caption{Exponents of the mass-luminosity relation of the form $L\sim M^{\alpha} \mu^{\beta}$ for homogeneous stars from \citet{Graefener_2011},
for a chemical composition corresponding to our zero-age main sequence models (labelled with the subscript ZAMS), and for pure helium stars of the corresponding 
metallicity (labelled with the subscript TAMS). Here, $\mu$ is the mean molecular weight of the stellar gas, which is computed as $1/\mu = 2X+0.75Y+0.5Z$, 
with $X$, $Y$, and $Z$ being the hydrogen, helium, and metal mass fraction, and assuming complete ionization.
}
    \label{fig_mld}
\end{figure}

We consider a mass-luminosity relation for chemically homogeneous stars as $L\sim M^{\alpha} \mu^{\beta}$,
where $\mu$ is the mean molecular weight of the stellar gas. The power-law exponent $\alpha$ thus describes the
slope of the ML-relation in the $\log M - \log L$-plane. According to \citet{Kippenhahn_1990},
we have $\alpha > 1$, and $\alpha \rightarrow 1$ for $M \rightarrow \infty$, due to the increasing radiation pressure 
for higher masses ($P_{\rm gas}/(P_{\rm gas} + P_{\rm rad}) \rightarrow 0$). The inequality $\alpha > 1$ implies
that, in the frame of simple opacity laws, massive main sequence stars will never exceed the Eddington 
limit. The limit of $\alpha \rightarrow 1$ implies that for sufficiently massive stars,
their hydrogen burning life time $\tau_H = E/\dot E$ becomes independent of mass, since their nuclear energy 
reservoir $E$ is proportional to their mass, just as their energy loss rate $\dot E = L$.  
Figure~\ref{fig_ml} shows that our most massive models reach $\alpha$-values very close to one.
We find the limiting stellar life time to be close to $1.9\,$Myr (cf. Table~B1). 

Figure~\ref{fig_mld} also shows that, at a given mass, the slope of the mass-luminosity relation is smaller
for the homogeneous hydrogen-free stars, compared to that for the zero-age main sequence stars because the helium stars are much more luminous and thus more radiation pressure dominated.
Furthermore, the power-law exponent $\beta$ also approaches one for the highest considered masses.  
While a $15\mso$ helium star is about 20 times as luminous as a $15\mso$ zero-age main sequence star
($\bar{\beta}(15\mso)\simeq 3.7$, and $\mu_{\rm tams}/\mu_{\rm zams}\simeq 2.23$), the corresponding
factor is only 2.4 at $500\mso$.

\subsection{Near the Eddington limit}
\label{sec3_inf}

In its general form, the Eddington limit is complex as it involves an appropriate mean of
the total of all opacity sources. This total opacity $\kappa$ may be a function of depth in the stellar
atmosphere; therefore, it is non-trivial to uniquely define the circumstances in which the star encounters its
Eddington limit, nor to assess the consequences such an encounter may have. 
The proximity of a star of mass $M$ and luminosity $L$ to the Eddington limit is usually expressed in 
terms of the Eddington factor
\begin{equation}
   \Gamma = \frac{\kappa \, L}{4\pi\,c\,G\,M} = \frac{\kappa\,\sigma\,T_{\rm eff}^{4}}{c\,g},
   \label{eq:Gamma}
\end{equation}
where the constants have their usual meaning and $T_{\rm eff}$ is the effective temperature and
$g$ the surface gravity.

Considering only the dominant contributor to the opacity, i.e. photon scattering on free electrons,
greatly simplifies the
concept of Eddington limit as electron scattering is a grey process and, for the hot stars considered here,
independent of depth in the atmosphere.  For a star of mass $M$ and luminosity $L$ we obtain
$\Gamma_{\rm e} = \kappa_{\rm e} L /(4 \pi c G M)$, where 
$\kappa_{\rm e} =\sigma_{\rm e} (1+X)$ is the opacity due to
Thomson scattering. It holds that $\Gamma > \Gamma_{\rm e}$ as $\kappa > \kappa_{\rm e}$.
Even in the most massive stars, the Eddington factor $\Gamma_{\rm e}$ will not exceed unity 
(see Sect.~\ref{sec3_ml}). Even so, since the mass-luminosity exponent $\alpha > 1$, main sequence stars
will get ever closer to the Eddington limit for electron scattering the higher their mass.


Considering the true Eddington factor, we must expect that the situation $\Gamma = 1$ is actually
achieved in very massive stars. 
Indeed, while $\kappa_e\simeq 0.34$
in our zero-age models, their true surface opacity is typically $\kappa_{\rm surface} \simeq 0.5$.
We can thus expect that $\Gamma = 1$ is achieved at $\Gamma_e \simeq 0.7$. 

\begin{figure*}[htbp]
    \centering
     \includegraphics[angle=-90,width=16cm]{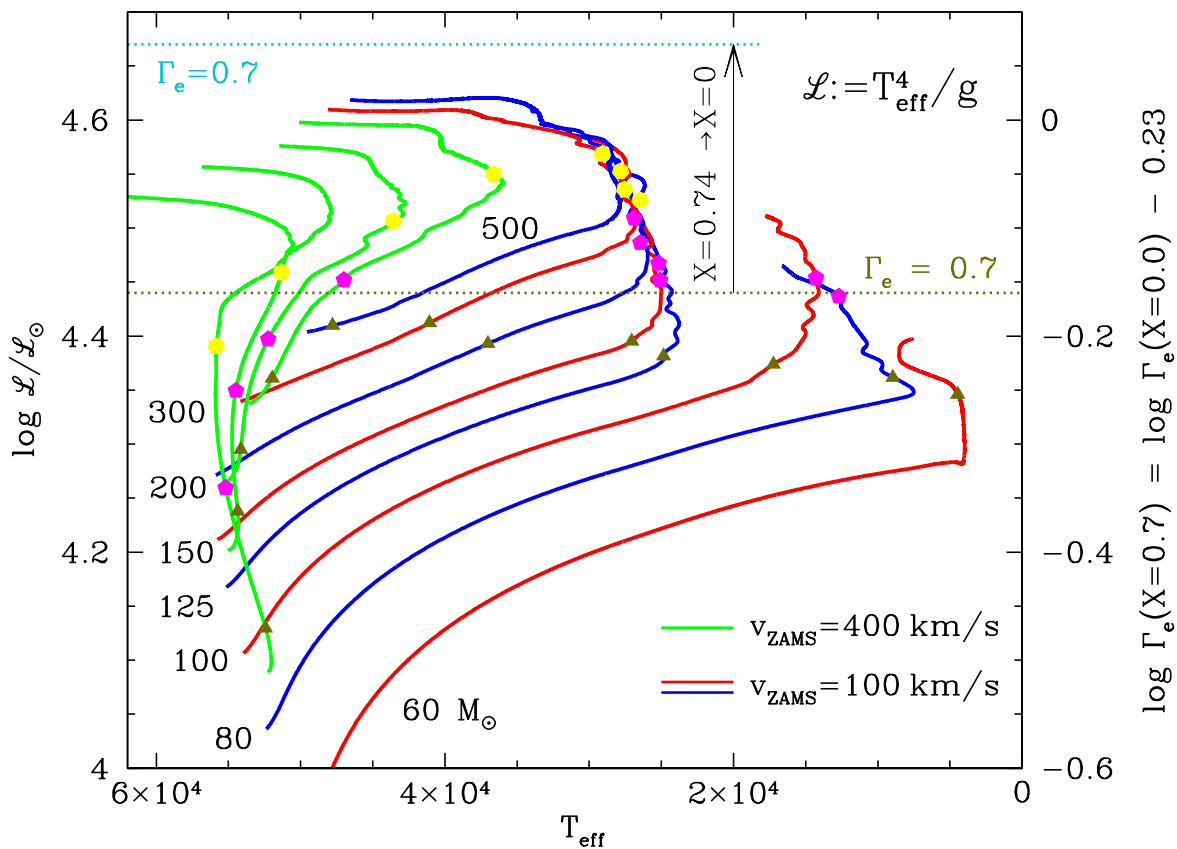}
    \caption{The evolutionary tracks in the $\mathscr L -  T_{\rm eff}$-diagram,
where  ${\mathscr L}=T_{\rm eff}^4/g$, for models initially rotating with 100\,km/s (for 60, 80, 100, 125, 150, 200,
300, and 500$\mso$; alternating red and blue lines) and with 400\,km/s (100, 150, 200, and 300$\mso$; green lines).
The right y-axis shows the Eddington factor for electron scattering opacity $\Gamma_e$, which is proportional
to $\mathscr L$, for a hydrogen mass fraction of $X=0.74$. For $X=0$, $\log \Gamma_e$ is larger by 0.23.
Triangles, pentagons, and heptagons mark the locations where the surface helium mass fraction
reaches 0.3, 0.5, and 0.7, respectively, for the presented evolutionary sequences.
The dotted horizontal lines marks the value of $\Gamma_e=0.7$ for $X=0.74$ (green) and $X=0$ (blue), approximately identifying the true
Eddington limit that cannot be exceeded (see text).
}
    \label{fig_L}
\end{figure*}

In Fig.~\ref{fig_L}, we plot the quantity ${\mathscr L} := T_{\rm eff}^4/g = c / 
(\kappa_{\rm e} \sigma) \,\Gamma_{\rm e}$
as a function of the effective temperature of selected evolutionary sequences.
We note that ${\mathscr L}$ is normalized to the solar value, 
$\log {\mathscr L}_{\odot} \simeq 10.61$ for convenience. 
For instance, for the initial composition of our stars we have
$\log {\mathscr L} / {\mathscr L}_{\odot} \simeq 4.6 + \log \Gamma_{\rm e}$.
The quantity ${\mathscr L}$ is proportional to $\Gamma_{\rm e}$ and can be derived
from spectroscopic observation of stars without knowledge of their distance
(cf. \citet{Langer_2014}).

Since for stars of constant mass 
${\mathscr L} \sim L $, the quantity ${\mathscr L}$ behaves in a similar way to the stellar
luminosity, and the evolutionary tracks in the ${\mathscr L}$-$T_{\rm eff}$-diagram 
partly resemble those in the Hertzsprung-Russell diagram. However, unlike the latter,
the ${\mathscr L}$-$T_{\rm eff}$-diagram has an impenetrable upper limit: the Eddington limit.
 
Comparing Fig.~\ref{fig_L} with Fig.~\ref{fig_hrd}, we see that while the tracks 
of the slow rotators in 
Fig.~\ref{fig_hrd} are almost horizontal, the corresponding tracks in Fig.~\ref{fig_L} are significantly
steeper. Their luminosities are nearly constant despite strong mass loss, because their increasing
mean molecular weight $\mu$ compensates for their decreasing mass in the mass-luminosity 
relation (Sect.~\ref{sec3_ml}). While it remains hidden in Fig.~\ref{fig_hrd},
Fig.~\ref{fig_L} reveals nicely that the mass loss drives these stars towards the Eddington
limit. The same is true for the rapidly rotating stars; their tracks move steeply upward in both diagrams.

In Fig.~\ref{fig_L}, we have drawn the horizontal line indicating $\Gamma_{\rm e} = 0.7$
for our initial chemical composition
--- which corresponds to $\log {\mathscr L} / {\mathscr L}_{\odot}\simeq 4.445$.
From the discussion above, we should expect that stellar models with unchanged surface
composition should not be found above this line (as long as their hydrogen and helium can be considered
fully ionized). This is confirmed by the triangles placed on the evolutionary tracks,
indicating the position at which a surface helium mass fraction of $Y=0.3$ is achieved,
and which are all found below $\Gamma_{\rm e} = 0.7$. Similarly, even the models with the most 
helium-enriched surfaces stay below the upper horizontal line for 
$\Gamma_{\rm e}(X=0) = 0.7$; in other words, the diagram shows that
it is not the electron-scattering Eddington limit which constrains the evolution of the most
massive stars, but the true Eddington limit. Furthermore, Fig.~\ref{fig_L} indicates
that the limiting electron-scattering Eddington factor --- which can be read off
from Fig.~\ref{fig_L} to be a little bit below $\Gamma_e = 0.7$ ---
is independent of the hydrogen/helium surface abundances.

\begin{figure}[htbp]
    \centering
     \includegraphics[angle=-90,width=8cm]{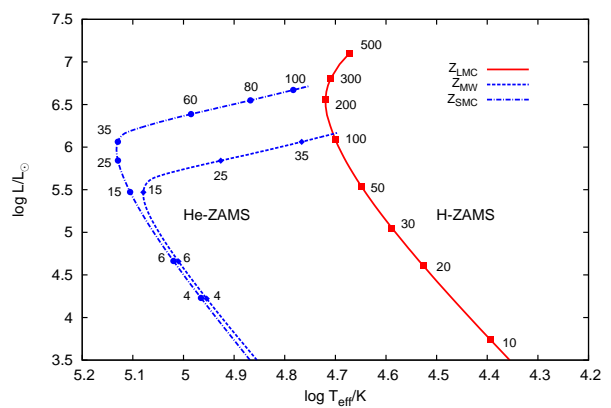}
    \caption{The zero-age main sequence of our non-rotating stellar models in the indicated mass range (solid red line), compared with zero-age main sequences helium star models for Milky Way (dashed blue line) and SMC (dash-dotted blue line) composition. Labelled dots along the lines imply masses in solar units. Owing to envelope inflation the helium main sequences bend towards cooler temperatures close to the Eddington limit.}
    \label{fig_inf}
\end{figure}

We find our models near the Eddington limit to undergo a significant envelope inflation.
This phenomenon has been described before by \citet{Kato_1986}  and \citet{Ishii_1999} for zero-age main sequence stars
of different metallicities, and by \citet{Ishii_1999, Petrovic_2006, Graefener_2012}
for Wolf-Rayet stars. Figure~\ref{fig_inf} shows the location of our zero-age main sequence stars, and of zero-age helium stars
with the same metallicity, in the HR-diagram. Both curves show a maximum effective temperature, which is reached at about
$30\mso$ (140\,000\,K) for the helium stars, and at $200\mso$ (54\,000\,K) for the hydrogen-rich stars. 
Above these masses, the effective temperature decreases again owing to the envelope inflation.

As seen in Fig.~\ref{fig_inf}, the line describing the helium stars bends sharply to cooler
temperatures at about 1\,000\,000$\lso$, and is even expected to cross the hydrogen-ZAMS
(see also \citet{Ishii_1999}), implying that more luminous helium stars are cooler
than equally bright zero-age main sequence stars. This has the consequence that stars undergoing chemically homogeneous
evolution above $\sim 125\mso$ evolve to cooler surface temperature (see Fig.~\ref{fig_hrd}).

The envelope inflation occurs because layers inside the stellar envelope reach or exceed the true Eddington limit.
The inflation begins before the critical value of $\Gamma_{\rm e} \simeq 0.7$ is reached, i.e. the electron-scattering Eddington factors are only $\Gamma_{\rm e} \simeq 0.31$ and $\Gamma_{\rm e}\simeq 0.41$
for the $15\mso$ helium star and $190\mso$ zero-age main sequence star, respectively, because of the
opacity peaks of iron and helium which can reach values of $1\,$cm$^2$/g or more.
While the subsurface layers with such opacities always turn convectively unstable, convection,
according to the mixing-length theory, is mostly very inefficient in transporting energy in these layers.
Consequently, the high radiation pressure pushes the overlying layers outwards, until the opacity
drops such that the true Eddington factor falls short of $\Gamma = 1$. 

Many of these models also develop density inversions near the layer with the maximum Eddington factor
\citep{Joss_1973}.
The corresponding inward directed gas pressure gradient then allows the star to retain layers
whose Eddington factor exceed the critical value of $\Gamma = 1$. While the stability of these structures
remains to be investigated, any instability which would tend to iron out the density inversion will likely
lead to a further inflation of the overlying layers. A further degree of complexity is added by the fact that
we find many models close to the Eddington limit to be unstable at least to radial pulsations (the only pulsation mode
which we can see with our one-dimensional hydrodynamic stellar evolution code), and that convective velocities close to
the sound speed may imply significant acoustic fluxes 
\citep{Goldreich_1990}.
A deeper investigation of these phenomena exceeds the scope
of the present paper and will be pursued elsewhere.

\section{Comparison with previous results}
\label{sec4}

\begin{figure}[htbp]
    \centering
     \includegraphics[angle=-90,width=8cm]{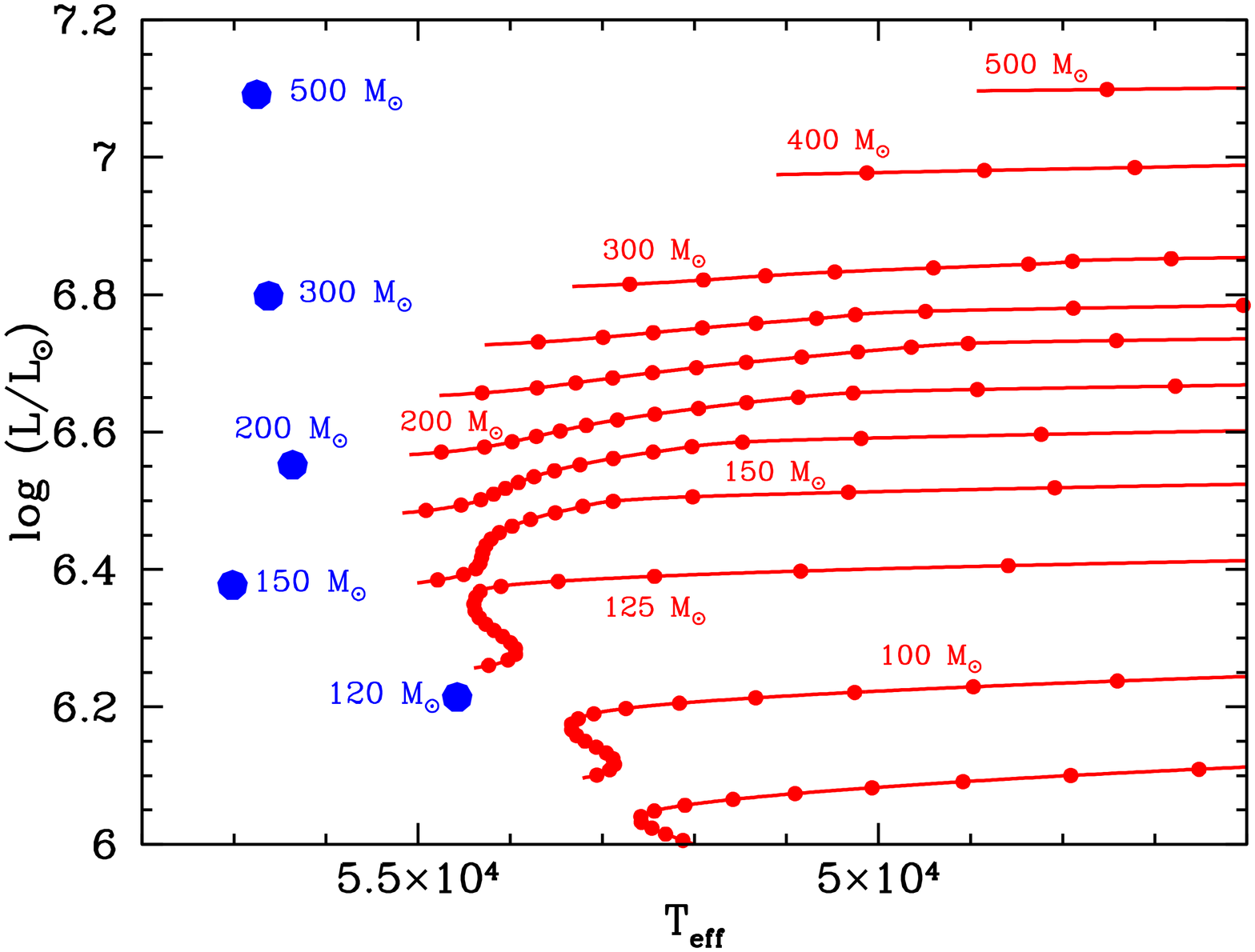}
    \caption{Comparison of the ZAMS position of the models from \citet{Yusof_2013} which include
rotation (blue dots) with our tracks for $\varv_{\rm rot,i}\simeq 300\,$km/s (red lines) in the HR diagram. 
We show tracks with initial masses of 80$\mso$, 100$\mso$, 125$\mso$, 150$\mso$, 175$\mso$, 200$\mso$,
230$\mso$, 260$\mso$, 300$\mso$, 400$\mso$, and 500$\mso$. The time interval between two dots on the
tracks corresponds to $10^5\,$yr.
}
    \label{fig_comp}
\end{figure}

\begin{figure}[htbp]
    \centering
     \includegraphics[angle=-90,width=8cm]{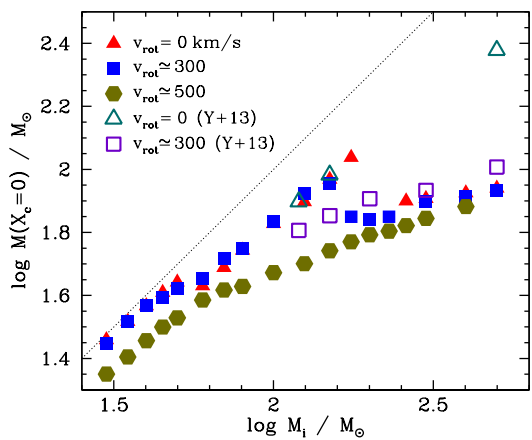}
    \caption{Stellar mass at the end of core hydrogen burning versus initial mass
for our sequences with initial rotational velocities of $v_{\rm rot,i}\simeq 0\,$km/s,
300\,km/s, and 500\,km/s (filled symbols). The open symbols correspond to the
LMC models of \citet{Yusof_2013} with and without rotation. The dotted straight line
corresponds to the location where the stellar mass at the end of core hydrogen burning
equals the initial mass, i.e. to negligible mass loss.
}
    \label{fig_mf}
\end{figure}

Recently, \citet{Yusof_2013} published a grid of stellar evolution models for very massive stars.
Eight of their evolutionary sequences were modelled using a similar initial composition to the one we used here, 
except that Yusof et al. used a scaled solar metal mix, while we take the measured LMC composition
where possible (cf. \citet{Brott_2011}). Yusof et al. present non-rotating models with initial masses
of 120$\mso$, 150$\mso$, and 500$\mso$ as well as rotating models at 120$\mso$, 150$\mso$,  200$\mso$,  
300$\mso$, and 500$\mso$. The rotating models of Yusof et al. had initial rotational velocities
of 400\,km/s or more (cf. their Table~2). 

\subsection{Subsurface convection and envelope inflation}

A major difference between the models of Yusof et al. and our models occurs because of the different treatment
of convection in the stellar envelopes. Starting at zero age, all our models contain subsurface convection zones  
which occur because of the opacity peaks of iron and helium (cf. \citet{Cantiello_2009}).
When using the standard mixing-length theory, where the mixing length is assumed to be proportional to the
pressure scale height, we find the convective energy transport in these zones to be mostly inefficient
because of the small local thermal timescales imposed by the low densities. As a consequence, we obtain a rich
envelope phenomenology as described in Sect.~\ref{sec3_inf}, including envelope inflation and density inversions.

Yusof et al. describe this situation as unphysical, and avoid it by assuming the convective mixing length 
to be proportional to the density scale height. As a consequence, when the star attempts to establish a density
inversion, the density scale height tends to infinity, which imposes the convective energy transport 
to be extremely efficient, and opacity peaks will have few consequences. While the existence of inflated envelopes
in nature remains to be shown (cf. \citet{Graefener_2012, Graefener_2013}),
we consider it likely, since the short thermal timescales of the envelopes of very massive stars
cannot be in accordance with efficient convective energy transport. While this will surely be studied in more
detail in the near future, we restrict ourselves here to pointing out the consequences of the different assumptions
on the convective efficiency.

Strong differences between the models of Yusof et al. and our models already occur at the zero-age main sequence.
As described in Sect.~\ref{sec3_trac}, our zero-age models (with our metallicity of $Z=0.0047$) reach a maximum 
effective temperature of about 56\,000\,K at $\sim 190\mso$, which agrees very well with the results of
\citet{Ishii_1999} for $Z=0.004$. In Fig.~\ref{fig_comp}, we compare the location of the rotating zero-age models of
Yusof et al. (according to their Table 2) with the early hydrogen burning evolution of our models with
an initial rotational velocity of $\sim 300\,$km/s. Above $\sim 120\mso$, the models of Yusof et al. are
hotter than ours, and the more so the larger the mass. 

This discrepancy is likely explained by the difference in the treatment of the envelope convection, since our treatment
yields a larger radius inflation the closer the star is to its Eddington limit (Sect.~\ref{sec3_inf}). 
At a given initial mass, the discrepancy between the models of Yusof et al. and our models is therefore expected to
increase with time at least initially, since the models increase their luminosity and decrease their mass
(cf. Fig.~\ref{fig_ml}), and thus their Eddington factors grow. The biggest consequence of this is that ---
although we use a very similar mass-loss prescription to that of Yusof et al. --- our mass-loss rates
are higher because our stars have cooler surfaces.  

\subsection{$M_{\sc ZAMS}$ versus $M_{\sc TAMS}$ relation}

Figure~\ref{fig_mf} shows that as a result of the different treatment of envelope convection in our models
and those of Yusof et al., our non-rotating models above $\sim 150\mso$ undergo a dramatic mass
loss during core hydrogen burning, leading to masses at core hydrogen exhaustion of the order of 
$80\mso$ even up to the highest initial mass of $500\mso$. Yusof et al. did not consider non-rotating models
between $150\mso$ and $500\mso$. However, their non-rotating
models below $150\mso$ and at $500\mso$
remain close to the diagonal in Fig.~\ref{fig_mf}. While our $500\mso$ model ends hydrogen
burning with a mass of $87\mso$, theirs does so with $239\mso$. We discuss the consequences for the final fate
of the stars in Sect.~\ref{sec4}. Remarkably, because of the low mass-luminosity exponent of the most
massive stars considered here, the core hydrogen burning life time of the $500\mso$ sequence of
Yusof et al. (1.9\,Myr) resembles ours, indicating that the life time of the most massive stars is not
strongly affected by the differences discussed above.

The non-rotating $500\mso$ sequence of Yusof et al. ends core hydrogen burning as a nearly chemically homogeneous
helium star with a surface helium mass fraction of 0.97. Our Fig.~\ref{fig_inf} above shows that 
at the considered metallicity, our helium star models start to have inflated envelopes above
$\sim 30\mso$ (see also \citet{Ishii_1999}, and \citet{Petrovic_2006}), and that
a $239\mso$ helium star model would be so enormously inflated that it would appear as a red supergiant,
if a stable envelope structure existed at all. This demonstrates that while the envelope physics used in
our models leads to pushing the outer layers of stars at the Eddington limit to large radii, this 
seems to occur to a much lesser degree in the models of Yusof et al., with the consequence of a smaller
mass-loss rate.

When looking at the Eddington factor for electron scattering $\Gamma_{\rm e}$, the models of Yusof et al. 
behave as our models do, i.e. they remain mostly lower than $\Gamma_{\rm e}=0.7$ (cf. Sect.~\ref{sec3_inf}).
However, their non-rotating $500\mso$ sequence is an exception, where $\Gamma_{\rm e}=0.82$ 
is reached at core hydrogen exhaustion. In comparison to the stellar wind calculations of
\citet{Vink_2011}, who found a strong mass-loss enhancement for stars
above $\Gamma_{\rm e}=0.7$, we would again expect that the mass at core hydrogen exhaustion of this sequence 
could be considerably lower than that found by Yusof et al.

Our models show the largest mass-loss rates during their Wolf-Rayet phase (cf. Sect.\,3.2). 
Nevertheless, it is not
the choice of the Wolf-Rayet mass-loss prescription which makes our non-rotating models lose more mass than 
the comparable models of Yusof et al. We find, for example, that for our slowly rotating 500$\mso$ models at the time of the maximum
mass-loss rate, the prescription by \citet{Nugis_2000} leads to a mass-loss rate for our stars which is two times larger than the one we use. 
We conclude that the high pre-Wolf-Rayet mass loss 
imposed by the inflated envelopes of our models plays a crucial role in explaining the smaller TAMS-mass
of our very massive slow rotators compared to those of Yusof et al. 
 
\subsection{Rotational mixing of helium} 
 
Another difference between the models of Yusof et al. and our models concerns the rotational mixing of helium.
As shown in Sect.~\ref{sec3_surf}, the mixing of helium in our models is practically absent below a threshold
rotational velocity, and nearly complete above the threshold. The mixed models suffer more mass loss since they 
become Wolf-Rayet stars earlier. As the threshold velocity decreases for higher masses,
this results in a bimodal behaviour of the total mass lost during core hydrogen burning (Fig.~\ref{fig_mf}). 
Our models with an initial rotational velocity of $\sim 300\,$km/s follow the non-rotating models
below $\sim 150\mso$, while they behave in a similar way to our fastest rotators for higher mass.
In the models of Yusof et al., the mixing of helium is gradually increased
for larger rotational velocities. Therefore, their rotating models end core hydrogen burning
with masses that exceed those of our homogeneously evolving models.

A more detailed comparison with the models of Yusof et al. is difficult because they considered only two rotational 
velocities, i.e. $\varv_{\rm rot,i} = 0\,$km/s and $\varv_{\rm rot,i}\simeq 400\,$km/s. This may also prevent 
a clear discrimination of their models from ours through observations, since 
\citet{Agudelo_2013} showed that only very few O stars are expected with
$\varv_{\rm rot,i} = 0\,$km/s and $\varv_{\rm rot,i}\simeq 400\,$km/s, while the majority rotate with
$\varv_{\rm rot,i} = 100 \dots 200\,$km/s. 


%
%

\section{Summary}
\label{sec5}
We present a detailed grid of stellar evolution models for single stars
with initial masses from 70 to 500\,$M_{\odot}$ and rotation rates up to 550 km/s.
We used the same physics and assumptions as \citet{Brott_2011} did for stars
in the mass range 5-60$\mso$, and this new grid therefore is an extension of their work.
The initial composition of our models corresponds to abundance measurements for massive stars in the Large Magellanic Cloud. 
We follow the evolution of the stellar models through their core hydrogen burning phase, with some  
of them computed well beyond this stage. 

Given the high fraction of close binaries in massive stars \citep{Sana_2012}, we cannot hope to obtain a complete picture
of massive star evolution from our models \citep{deMink_2014}, which is true in particular for the most massive stars
\citep{Schneider_2014}. On the other hand, our rotating models may constitute a fair approximation 
for those mass gainers and mergers in close binaries which rejuvenate after the binary interaction
\citep{Braun_1995}, even though their age, their detailed surface abundances, and 
their spin may still be peculiar \citep{deMink_2013}. In addition, since most mass donors
in post-interaction close binaries either remain unobserved \citep{deMink_2014}
or show extreme surface enrichments \citep{Langer_2012}, and as \citet{Sana_2012,Sana_2013b} suggest that 30-50\% of the massive stars in 30 Dor appear not to have a close companion, a comparison of our single-star models with observed very massive stars will still be meaningful. 

We find that stellar rotation can influence the evolution of our models significantly, mostly through mixing of
helium. We find two threshold initial rotational velocities, which both decrease with increasing initial mass,
one below which the rotational mixing of helium is negligible, and a second one above which the
models undergo quasi-chemically homogeneous evolution. For initial rotational velocities in between the threshold values,
the models start chemically homogeneously and develop helium-enriched surfaces, but transit to normal evolution as a result of spin-down. 

Above an initial mass of $\sim 160\mso$, we find that quasi-chemically 
homogeneous evolution can also be achieved through
mass loss, which in fact ensures that all our more massive models end core-hydrogen
burning as nearly pure helium stars, independent of their initial rotation rate. We find that single stars
with initial rotational velocities below 300\,km/s need to be more massive than $\sim 100\mso$ in order to
achieve a surface helium mass fraction above 35\% (cf. Fig.~\ref{fig_NY}). 

Significant mixing of trace elements, including nitrogen and boron, is found well below the threshold velocities 
mentioned above. Because of the high core temperatures in our very massive models, the NeNa- and MgAl-cycles
are also activated and lead to surface abundance changes of the involved isotopes.
A tripling of the surface nitrogen abundance occurs in our models above an initial rotational velocity
of 150\,km/s and a mass of $\sim 25\mso$, while a doubling at this rotation rate occurs even down to 
$6\mso$ (Fig.~\ref{fig_NH2}). Because of the strong mass loss and large convective core masses, the surfaces of our models
above $\sim 100\mso$ become nitrogen-rich irrespective of their rotation rate. 

Our zero-age models above $\sim 150\mso$, as well as more evolved models down to $\sim 40\mso$ show significant
envelope inflation because the true Eddington factor inside the stellar envelope approaches or even exceeds
the value of unity, while the electron-scattering Eddington factor remains below or near 
a value of $\Gamma_{\rm e}=0.7$ (cf. Sect.~\ref{sec3_inf}). 
Consequently, our zero-age main sequence models show a maximum surface temperature of 56\,000\,K
at about $180\mso$, with lower values for higher masses (Fig.~\ref{fig_hrd}).

During core hydrogen burning, the envelope inflation drives our slow rotators in the mass range
$50-100\mso$ to effective temperatures near or below 10\,000\,K.
At these cool temperatures, partial hydrogen recombination leads to high opacities, which in
turn produces true Eddington factors well above one in the envelopes of these models.
While a detailed study of this phenomenon remains to be done, we speculate that this feature is
related to the outbursts of luminous blue variables.  

The inflation and the corresponding decrease of the
surface temperature lead to enhanced mass loss compared to the models of \citet{Yusof_2013}.
The enhanced mass loss has strong consequences for the final fate of very massive stars. 
In this context it is important to emphasize the metallicity dependence of inflation \citep{Ishii_1999}.
The reduced masses (Fig.~\ref{fig_mf}) exclude
very massive stars of the considered metallicity as progenitors of pair-instability supernovae.
This is in agreement with \citet{Langer_2007}, who demonstrated that models with the same
envelope physics as used here may produce pair instability supernovae at a metallicity
of $Z_{\odot}/20$ \citep{Kozyreva_2014}.  
This is in contrast to the finding of Yusof et al.,
who suggest that slow rotators above $\sim 300\mso$ end up with CO-cores above $60\mso$
at a metallicity of that of the LMC.

A further consequence of the strong mass loss of our inflated models is their distinct spin-down
during core hydrogen burning (Fig.~\ref{fig_velo}).
This prevents all our models above $\sim 60\mso$ from producing a long-duration gamma-ray burst
upon collapse. From the models considered here and including those from \citet{Brott_2011},
the highest chance to produce long-duration gamma-ray burst at LMC metallicity from the chemically homogeneous evolution
scenario \citep{Yoon_2005, Woosley_2006} may occur in the mass range $20-30\mso$ \citep{Langer_2012}.

The VLT-FLAMES Tarantula Survey \citep{Evans_2011} provides one of the major motivations for this study.
Our models will be extensively compared with these and other observations in the near future.
To this end, we provide the main outputs of our stellar evolution models as electronic tables (stored at CDS)
in the same format as those of \citet{Brott_2011}. Additionally, we provide sets of isochrones in Appendix~\ref{app1}.

\begin{acknowledgements} 
We want to thank Durand D'Souza for fruitful discussions, and Jes\'us Ma\'iz Apell\'aniz, Matteo Cantiello, Joachim Puls, Jon Sundqvist and Paco Najarro for useful comments.
SdM acknowledges support for this work by NASA through an Einstein Fellowship grant, PF3-140105.
We are grateful to the referee for improving this paper.
\end{acknowledgements}

\begin{appendix}

\newpage

\section{Isochrones}
\label{app1}

The evolution of our rapidly rotating stellar models is influenced significantly by rotation (see Sect.~\ref{sec3_trac}). Therefore, 
isochrones of rotating models calculated for the same age differ from isochrones of non-rotating models. 

\begin{figure}[htbp]
    \centering
     \includegraphics[angle=-90,width=8.5cm]{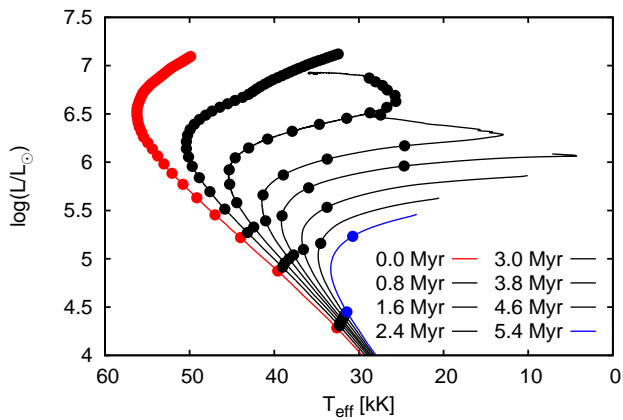}
    \caption{Isochrones of non-rotating stellar evolution models in the mass range 10 - 500\,$M_{\odot}$ are depicted 
for ages up to 5.4 Myr. The different line colours explained in the figure key indicate the age of every isochrone. Models in 10\,$M_{\odot}$-steps are highlighted by filled circles.}
    \label{fig_iso_age}
\end{figure} 

Figure~\ref{fig_iso_age} depicts isochrones of non-rotating models for ages up to 5.4 Myr in the HR diagram. Eight 
different isochrones are shown with age steps of 0.8 Myr. The core hydrogen burning stellar evolution models presented in Sect.~\ref{sec3} 
are used to generate the isochrones. Because more massive stellar models have shorter lifetimes, 
older isochrones terminate at the less massive model at the upper end of the track. 

\begin{figure*}[htbp]
    \centering
     \includegraphics[width=15.5cm]{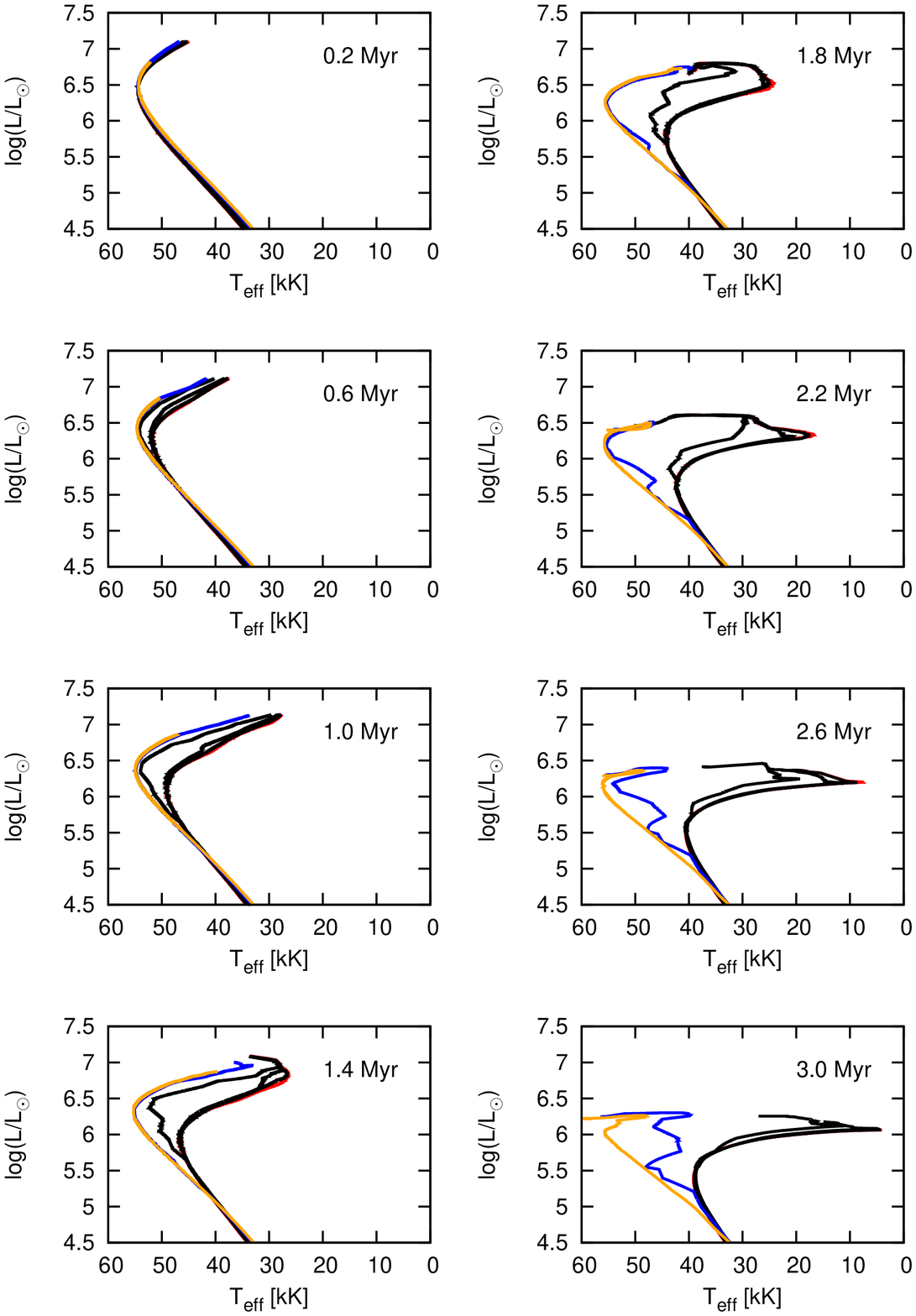}
    \caption{Several isochrones of different ages and rotating stellar evolution models are shown in the HR diagram. The surface rotational velocity at the ZAMS is chosen in steps of 100\,km/s from non-rotating to 400\,km/s and additionally 450\,km/s. Three initial surface rotational velocities are highlighted in particular. The isochrones of non-rotating models are shown in red, 400\,km/s in blue, and 450\,km/s in orange.}
    \label{fig_iso_vrot}
\end{figure*}

\begin{figure*}[htbp]
    \centering
     \includegraphics[width=15.5cm]{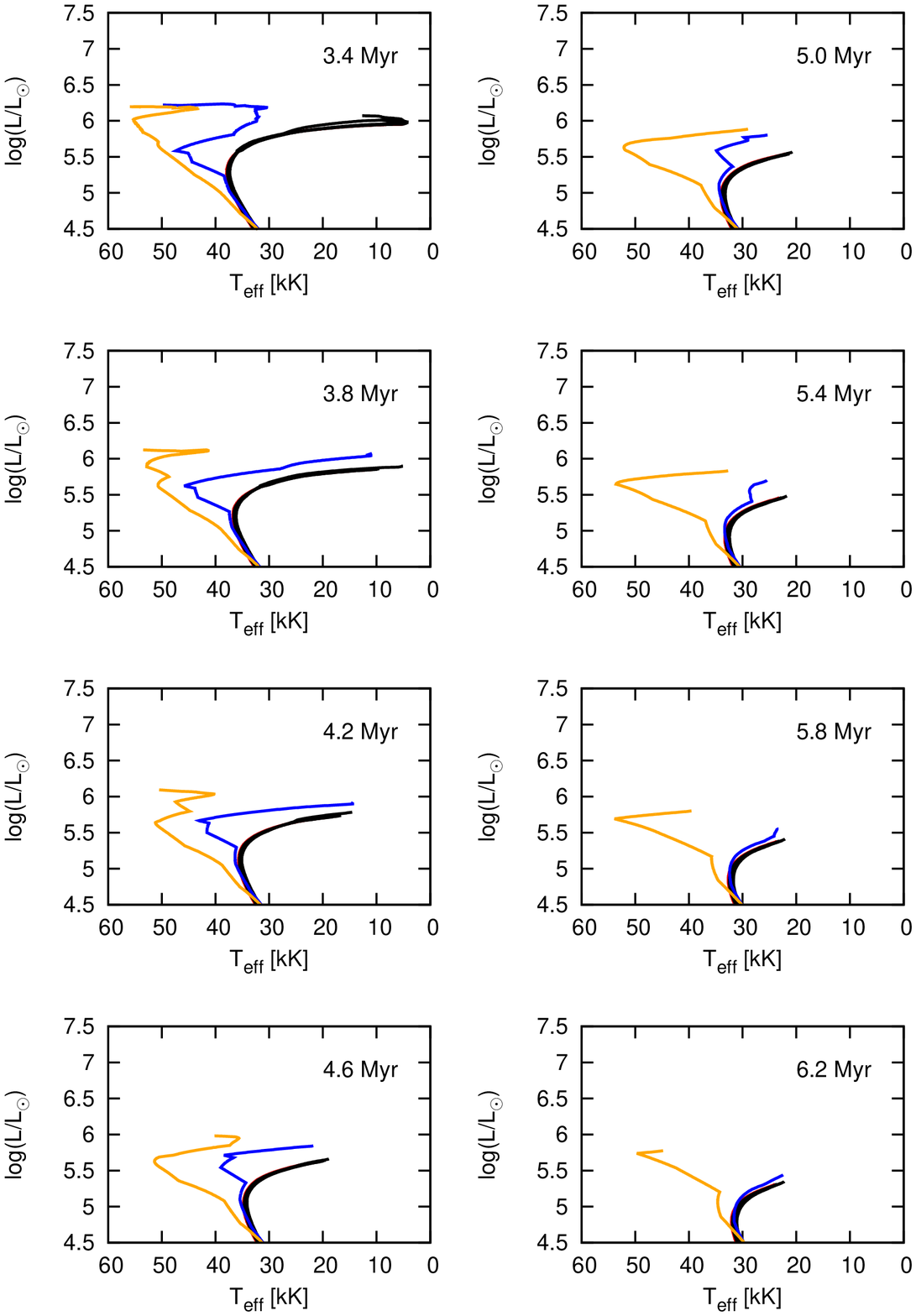}
    \caption{Several isochrones of different ages and rotating stellar evolution models are shown in the HR diagram. The surface rotational velocity at the ZAMS is chosen in steps of 100\,km/s from non-rotating to 400\,km/s and additionally 450\,km/s. Three initial surface rotational velocities are highlighted. The isochrones of non-rotating models are shown in red, 400\,km/s in blue, and 450\,km/s in orange.}
    \label{fig_iso_vrot2}
\end{figure*}

For a given initial composition and age, isochrones span an area in the HR diagram when different initial surface rotational 
velocities are considered simultaneously \citep{Brott_2011}. Isochrones of 
16 different ages from 0.2 to 6.2 Myr for
rotating stellar models are shown in 
Fig.~\ref{fig_iso_vrot} and ~\ref{fig_iso_vrot2} for different ages.
Switching from inhomogeneous to chemically homogeneous evolution, the isochrone is located at 
higher effective temperatures and luminosities than for the non-chemically homogeneous evolution.

The more massive a stellar model is, the earlier it reaches the point in the HR diagram 
where the stellar evolution track starts to evolve blueward. The related turn in the 
isochrones in Fig.~\ref{fig_iso_vrot} is first visible for 1.4 Myr (bottom left panel). The minimum effective 
temperature depends on the surface rotational velocity at the ZAMS, which is reflected by the isochrones 
of different rotation rates. The 2.2 Myr and 2.6 Myr isochrones show that the more massive a stellar model is, 
the earlier the \citet{Hamann_1995} mass-loss rate is applied to the stellar evolution calculation. 
The isochrones therefore show a decrease in luminosity for the most massive models. 

\begin{figure}[htbp]
    \centering
     \includegraphics[angle=-90,width=8.5cm]{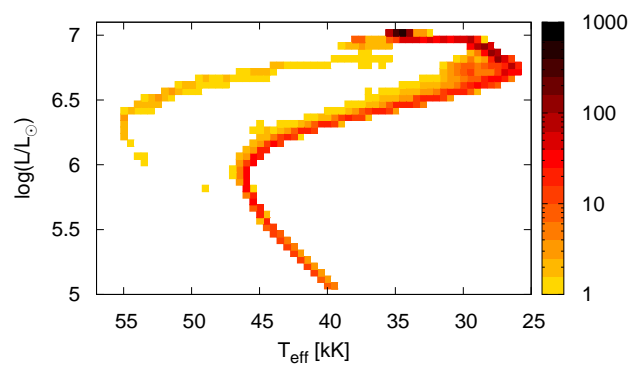}
    \caption{{\sc starmaker} population-synthesis calculation for 1.5\,Myr (and other parameters as explained in Table~\ref{tab_popsynpara}). The luminosity is depicted as a function of the effective temperature. The colour coding explained at the right bar indicates the number of stars within one pixel of 500\,K and $\log (L/L_{\odot})$=0.05.}
    \label{fig_iso_pop}
\end{figure}

Figure \ref{fig_iso_pop} depicts a population synthesis of stars with the age and distributions of mass and surface 
rotational velocity given in Table~\ref{tab_popsynpara}. The calculation was done using the code {\sc Starmaker} 
\citep{Brott_2011b}, using the parameters listed in Table~A1.

\begin{table}[h]
\caption{Parameters used in our population synthesis calculation (cf. Fig.~\ref{fig_iso_pop}).}
\label{tab_popsynpara}
\centering
\begin{tabular}{l l}
\hline\hline
parameter &  \\
\hline
age & 1.5 Myr\\
velocity distribution & Gaussian distribution \citep{Brott_2011b} \\
& ($\sigma=141\,$km/s, $\mu$=100\,km/s)\\
velocity range & 0 -- 500\,km/s\\
mass distribution & uniform distribution\\
mass range & 10 -- 500\,$M_{\odot}$\\
\hline
\end{tabular}
\end{table}

It can be seen that the randomly drawn stellar models with the same age do not lie on one line, but instead 
spread over a certain area in the HR diagram. The initial mass distribution determines the model density along the line 
corresponding to the isochrone of non-rotating stellar models of this age. We choose a flat mass distribution 
for the simulation to have a better view of the behaviour of the most massive stars. 
The velocity distribution on the other hand shapes the stellar model density as a function of the effective temperature for
(roughly) constant luminosity. The colour coding indicates the number of stars within one pixel of 500\,K and $\log (L/L_{\odot})$=0.05. 

Figure~\ref{fig_iso_pop} shows similar information as discussed in Fig.~\ref{fig_iso_vrot}. Additionally, it gives the 
probability of observing a star for the conditions given in Table~\ref{tab_popsynpara} depending on the surface rotational 
velocity distribution.

Isochrones for slow rotating stellar evolution models that undergo non-chemically homogeneous evolution do not differ significantly 
in the HR diagram. The probability of observing a star along these isochrones is highest. The change from non-chemically 
homogeneous to quasi-chemically homogeneous evolution occurs at a small velocity range related to a significant change 
of the position in the HR diagram.Therefore, the probability of observing a star between isochrones of slow-rotating and 
rapidly rotating models is small. Again, isochrones of rapidly rotating stellar evolution models are close in the HR diagram, leading to a higher probability of observation.

\section{Summary of data}
\label{app2}

\begin{table*}[h]
\caption{Summary of important stellar characteristics.}
\label{tab_data}
\begin{sideways}
\centering
\begin{tabular}{r r r r r r r r r r r r r r r r r}
\hline\hline
$M_\mathrm{i}$ & $v_{\mathrm{rot,i}}$ & $\tau_{\mathrm{MS}}$ & $\log L_{\mathrm{i}}/L_{\odot}$ & $T_{\mathrm{eff,i}}$ & $M_\mathrm{conv,i}$ & $M_\mathrm{over,i}$ & $M_\mathrm{conv,0.2}$ & $M_\mathrm{over,0.2}$ & $M_{\mathrm{f}}$ & $v_{\mathrm{rot,f}}$ & $Y_{\mathrm{s,f}}$ & $Y_{\mathrm{c,f}}$ & $\log L_{\mathrm{f}}/L_{\odot}$ & $T_{\mathrm{eff,f}}$ & $j$ \\
$M_{\odot}$ & km/s & Myr & & K & $M_{\odot}$ & $M_{\odot}$ & $M_{\odot}$ & $M_{\odot}$ & $M_{\odot}$ & km/s & & & & K & 10$^{15}$ cm$^2$/s \\
\hline
60  & 0   & 3.373 & 5.704 & 50000 & 42.6 & - & 30.7 & - & 43  & 0   & 0.393 & 0.995 & 6.022 &  6600 & 0 \\
60  & 220 & 3.407 & 5.701 & 49500 & 43.8 & - & 30.7 & - & 42  & 0   & 0.398 & 0.995 & 6.023 &  8800 & 89\\
60  & 273 & 3.464 & 5.699 & 49200 & 43.6 & - & 30.8 & - & 42  & 0   & 0.384 & 0.995 & 6.023 &  5600 & 89\\
60  & 326 & 3.619 & 5.698 & 48900 & 43.5 & 3.0 & 32.4 & 2.6 & 45  & 1   & 0.418 & 0.995 & 6.056 & 14200 & 85\\
60  & 379 & 3.676 & 5.694 & 48600 & 43.4 & 3.1 & 32.7 & 2.6 & 45  & 1   & 0.412 & 0.995 & 6.061 & 14800 & 84\\
60  & 432 & 4.038 & 5.691 & 48300 & 43.3 & 3.1 & 38.6 & 2.1 & 45  & 518 & 0.677 & 0.995 & 6.133 & 38900 & 237\\
60  & 532 & 4.167 & 5.684 & 47600 & 43.0 & 3.2 & 40.4 & 1.6 & 38  & 383 & 0.865 & 0.995 & 6.098 & 69600 & 212\\
100 & 0   & 2.648 & 6.102 & 54100 & 80.6 & 3.8 & 57.2 & 3.8 & 68  & 0   & 0.636 & 0.993 & 6.336 & 17600 & 0 \\
100 & 217 & 2.692 & 6.099 & 53700 & 82.1 & 3.6 & 57.4 & 3.7 & 66  & 1   & 0.674 & 0.991 & 6.331 & 22200 & 98\\
100 & 271 & 2.740 & 6.098 & 53500 & 81.9 & 3.5 & 58.2 & 3.7 & 65  & 12  & 0.722 & 0.993 & 6.337 & 27800 & 97\\
100 & 323 & 2.817 & 6.096 & 53200 & 81.7 & 4.3 & 60.1 & 3.5 & 67  & 7   & 0.720 & 0.993 & 6.354 & 26400 & 105\\
100 & 375 & 2.973 & 6.094 & 53000 & 81.6 & 3.5 & 66.3 & 2.6 & 61  & 64  & 0.929 & 0.991 & 6.345 & 47700 & 130\\
100 & 426 & 3.016 & 6.093 & 52700 & 81.4 & 3.5 & 66.3 & 2.2 & 50  & 39  & 0.974 & 0.993 & 6.236 & 58900 & 88\\
100 & 524 & 3.065 & 6.088 & 52100 & 81.0 & 3.7 & 66.1 & 1.8 & 46  & 64  & 0.975 & 0.995 & 6.213 & 73000 & 83\\
150 & 0   & 2.313 & 6.382 & 56000 & 131.6 & 3.7 & 91.3 & 4.4 & 94  & 0   & 0.839 & 0.975 & 6.553 & 30200 & 0 \\
150 & 216 & 2.339 & 6.381 & 55600 & 131.3 & 4.1 & 91.5 & 4.4 & 94  & 4   & 0.836 & 0.973 & 6.553 & 30200 & 121\\
150 & 268 & 2.417 & 6.380 & 55400 & 131.1 & 3.7 & 95.4 & 4.2 & 99  & 11  & 0.821 & 0.966 & 6.573 & 28900 & 144\\
150 & 320 & 2.465 & 6.378 & 55200 & 130.9 & 3.8 & 97.9 & 3.5 & 91  & 17  & 0.931 & 0.966 & 6.550 & 41100 & 139\\
150 & 371 & 2.544 & 6.377 & 55000 & 130.7 & 3.9 & 98.0 & 2.3  & 57  & 15  & 0.978 & 0.991 & 6.313 & 55000 & 47\\
150 & 421 & 2.557 & 6.376 & 54700 & 130.6 & 3.8 & 96.8 & 2.1 & 55  & 17  & 0.979 & 0.993 & 6.300 & 56400 & 48\\  
150 & 514 & 2.579 & 6.372 & 54200 & 130.2 & 3.9 & 95.3 & 2.1 & 55  & 17  & 0.977 & 0.991 & 6.293 & 55600 & 51\\
200 & 0   & 2.140 & 6.567 & 56200 & 181.5 & 3.9 & 126.1 & 4.7 & 131 & 0   & 0.809 & 0.916 & 6.698 & 30700 & 0 \\
200 & 214 & 2.155 & 6.566 & 55800 & 181.1 & 4.0 & 125.4 & 4.7 & 126 & 9   & 0.843 & 0.927 & 6.693 & 32300 & 159\\
200 & 267 & 2.185 & 6.565 & 55600 & 180.9 & 3.9 & 126.7 & 5.6 & 133 & 8   & 0.792 & 0.912 & 6.700 & 30300 & 190\\
200 & 318 & 2.255 & 6.564 & 55400 & 180.7 & 4.0 & 130.2 & 3.8 & 70  & 2   & 0.986 & 0.995 & 6.436 & 51100 & 37\\
200 & 367 & 2.291 & 6.563 & 55300 & 180.6 & 3.9 & 129.1 & 2.8 & 64  & 3   & 0.985 & 0.995 & 6.383 & 53800 & 37\\
200 & 415 & 2.302 & 6.562 & 55100 & 180.4 & 4.1 & 127.6 & 2.3 & 65  & 2   & 0.980 & 0.991 & 6.387 & 50900 & 41\\
200 & 502 & 2.334 & 6.560 & 54700 & 179.8 & 4.1 & 121.7 & 2.1 & 62  & 6   & 0.979 & 0.991 & 6.362 & 52400 & 43\\
300 & 0   & 1.979 & 6.811 & 54500 & 281.1 & 4.7 & 195.3 & 4.7 & 82  & 0   & 0.988 & 0.993 & 6.519 & 46800 & 0 \\
300 & 213 & 2.003 & 6.810 & 54300 & 280.6 & 3.8 & 193.4 & 4.9 & 80  & 1   & 0.988 & 0.993 & 6.509 & 47300 & 17\\
300 & 264 & 2.023 & 6.810 & 54200 & 280.6 & 3.8 & 193.7 & 4.7 & 80  & 1   & 0.988 & 0.993 & 6.507 & 47600 & 17\\
300 & 313 & 2.037 & 6.809 & 54100 & 280.6 & 3.8 & 193.7 & 4.7 & 79  & 1   & 0.987 & 0.995 & 6.505 & 48100 & 18\\
300 & 359 & 2.062 & 6.808 & 54000 & 280.2 & 4.0 & 191.8 & 3.2 & 75  & 1   & 0.987 & 0.993 & 6.473 & 49500 & 19\\
300 & 403 & 2.076 & 6.807 & 53800 & 280.2 & 4.0 & 187.0 & 1.7 & 73  & 1   & 0.986 & 0.993 & 6.459 & 49800 & 21\\  
300 & 475 & 2.099 & 6.806 & 53600 & 279.8 & 3.9 & 173.6 & 2.0 & 70  & 2   & 0.987 & 0.995 & 6.447 & 58900 & 24\\
500 & 0   & 1.855 & 7.097 & 49800 & 480.3 & 4.3 & 289.0 & 2.7 & 88  & 0   & 0.989 & 0.993 & 6.563 & 44900 & 0 \\
500 & 210 & 1.885 & 7.096 & 49700 & 480.3 & 4.3 & 283.1 & 2.4 & 86  & 0   & 0.988 & 0.993 & 6.551 & 45300 & 7\\
500 & 259 & 1.883 & 7.096 & 49700 & 479.9 & 4.7 & 282.9 & 2.1 & 86  & 0   & 0.988 & 0.993 & 6.550 & 45700 & 7\\
500 & 304 & 1.894 & 7.096 & 49600 & 478.6 & 4.9 & 280.8 & 2.3 & 86  & 0   & 0.988 & 0.993 & 6.549 & 45200 & 7\\
500 & 344 & 1.900 & 7.096 & 49500 & 476.8 & 4.7 & 359.1 & 7.7 & 400 & 345 & 0.534 & 0.659 & 7.134 & 29200 & 1077\\
500 & 379 & 1.922 & 7.095 & 49500 & 476.8 & 4.7 & 360.1 & 6.9 & 398 & 342 & 0.546 & 0.652 & 7.131 & 31500 & 1201\\
500 & 420 & 1.936 & 7.095 & 49400 & 476.3 & 4.8 & 354.7 & 5.0 & 380 & 154 & 0.609 & 0.676 & 7.129 & 33100 & 1051\\
\hline
\end{tabular}
\end{sideways}
\tablefoot{From left to right: Mass and surface rotational velocity at the ZAMS (index i); main-sequence lifetime; luminosity and effective temperature at the ZAMS; mass of the convective core and the overshoot zone at the first calculated model after the ZAMS and when hydrogen in the core drops below 0.2; mass, surface rotational velocity, helium mass fraction at the surface and in the core, luminosity, effective temperature; and surface specific angular momentum at the end of the main sequence evolution (index f). Several stellar evolution models have not been calculated until the TAMS. In these cases, the main sequence lifetime is extrapolated until the exhaustion of hydrogen in the core and the final values in Cols. 6 -- 16 correspond to the last calculated values. Dashes in the table indicate data which is not available.}
\end{table*}

\end{appendix}

\bibliographystyle{aa}
\bibliography{literatur}

\begin{thebibliography}{78}
\expandafter\ifx\csname natexlab\endcsname\relax\def\natexlab#1{#1}\fi

\bibitem[{{Bestenlehner} {et~al.}(2014){Bestenlehner}, {Gr{\"a}fener}, {Vink},
  {Najarro}, {de Koter}, {Sana}, {Evans}, {Crowther}, {H{\'e}nault-Brunet},
  {Herrero}, {Langer}, {Schneider}, {Sim{\'o}n-D{\'{\i}}az}, {Taylor}, \&
  {Walborn}}]{Bestenlehner_2014}
{Bestenlehner}, J.~M., {Gr{\"a}fener}, G., {Vink}, J.~S., {et~al.} 2014, \aap,
  570, A38

\bibitem[{{B{\"o}hm-Vitense}(1958)}]{Vitense_1958}
{B{\"o}hm-Vitense}, E. 1958, \zap, 46, 108

\bibitem[{{Bouret}(2004)}]{Bouret_2004}
{Bouret}, J.-C. 2004, in EAS Publications Series, Vol.~13, EAS Publications
  Series, ed. M.~{Heydari-Malayeri}, P.~{Stee}, \& J.-P. {Zahn}, 271--291

\bibitem[{{Bouret} {et~al.}(2012){Bouret}, {Hillier}, {Lanz}, \&
  {Fullerton}}]{Bouret_2012}
{Bouret}, J.-C., {Hillier}, D.~J., {Lanz}, T., \& {Fullerton}, A.~W. 2012,
  \aap, 544, A67

\bibitem[{{Bouret} {et~al.}(2005){Bouret}, {Lanz}, \& {Hillier}}]{Bouret_2005}
{Bouret}, J.-C., {Lanz}, T., \& {Hillier}, D.~J. 2005, \aap, 438, 301

\bibitem[{{Braun} \& {Langer}(1995)}]{Braun_1995}
{Braun}, H. \& {Langer}, N. 1995, \aap, 297, 483

\bibitem[{{Bresolin} {et~al.}(2008){Bresolin}, {Crowther}, \&
  {Puls}}]{Bresolin_2008}
{Bresolin}, F., {Crowther}, P.~A., \& {Puls}, J., eds. 2008, IAU Symposium,
  Vol. 250, {Massive Stars as Cosmic Engines}

\bibitem[{{Brott} {et~al.}(2011{\natexlab{a}}){Brott}, {de Mink}, {Cantiello},
  {Langer}, {de Koter}, {Evans}, {Hunter}, {Trundle}, \& {Vink}}]{Brott_2011}
{Brott}, I., {de Mink}, S.~E., {Cantiello}, M., {et~al.} 2011{\natexlab{a}},
  \aap, 530, A115

\bibitem[{{Brott} {et~al.}(2011{\natexlab{b}}){Brott}, {Evans}, {Hunter}, {de
  Koter}, {Langer}, {Dufton}, {Cantiello}, {Trundle}, {Lennon}, {de Mink},
  {Yoon}, \& {Anders}}]{Brott_2011b}
{Brott}, I., {Evans}, C.~J., {Hunter}, I., {et~al.} 2011{\natexlab{b}}, \aap,
  530, A116

\bibitem[{{Cantiello} {et~al.}(2009){Cantiello}, {Langer}, {Brott}, {de Koter},
  {Shore}, {Vink}, {Voegler}, {Lennon}, \& {Yoon}}]{Cantiello_2009}
{Cantiello}, M., {Langer}, N., {Brott}, I., {et~al.} 2009, \aap, 499, 279

\bibitem[{{Chieffi} \& {Limongi}(2013)}]{Chieffi_2013}
{Chieffi}, A. \& {Limongi}, M. 2013, \apj, 764, 21

\bibitem[{{Crowther} {et~al.}(2010){Crowther}, {Schnurr}, {Hirschi}, {Yusof},
  {Parker}, {Goodwin}, \& {Kassim}}]{Crowther_2010}
{Crowther}, P.~A., {Schnurr}, O., {Hirschi}, R., {et~al.} 2010, MNRAS, 408, 731

\bibitem[{{de Mink} {et~al.}(2013){de Mink}, {Langer}, {Izzard}, {Sana}, \& {de
  Koter}}]{deMink_2013}
{de Mink}, S.~E., {Langer}, N., {Izzard}, R.~G., {Sana}, H., \& {de Koter}, A.
  2013, \apj, 764, 166

\bibitem[{{de Mink} {et~al.}(2014){de Mink}, {Sana}, {Langer}, {Izzard}, \&
  {Schneider}}]{deMink_2014}
{de Mink}, S.~E., {Sana}, H., {Langer}, N., {Izzard}, R.~G., \& {Schneider},
  F.~R.~N. 2014, \apj, 782, 7

\bibitem[{{Evans} {et~al.}(2006){Evans}, {Lennon}, {Smartt}, \&
  {Trundle}}]{Evans_2006}
{Evans}, C.~J., {Lennon}, D.~J., {Smartt}, S.~J., \& {Trundle}, C. 2006, \aap,
  456, 623

\bibitem[{{Evans} {et~al.}(2005){Evans}, {Smartt}, {Lee}, {Lennon}, {Kaufer},
  {Dufton}, {Trundle}, {Herrero}, {Sim{\'o}n-D{\'{\i}}az}, {de Koter},
  {Hamann}, {Hendry}, {Hunter}, {Irwin}, {Korn}, {Kudritzki}, {Langer},
  {Mokiem}, {Najarro}, {Pauldrach}, {Przybilla}, {Puls}, {Ryans}, {Urbaneja},
  {Venn}, \& {Villamariz}}]{Evans_2005}
{Evans}, C.~J., {Smartt}, S.~J., {Lee}, J.-K., {et~al.} 2005, \aap, 437, 467

\bibitem[{{Evans} {et~al.}(2011){Evans}, {Taylor}, {H{\'e}nault-Brunet},
  {Sana}, {de Koter}, {Sim{\'o}n-D{\'{\i}}az}, {Carraro}, {Bagnoli}, {Bastian},
  {Bestenlehner}, {Bonanos}, {Bressert}, {Brott}, {Campbell}, {Cantiello},
  {Clark}, {Costa}, {Crowther}, {de Mink}, {Doran}, {Dufton}, {Dunstall},
  {Friedrich}, {Garcia}, {Gieles}, {Gr{\"a}fener}, {Herrero}, {Howarth},
  {Izzard}, {Langer}, {Lennon}, {Ma{\'{\i}}z Apell{\'a}niz}, {Markova},
  {Najarro}, {Puls}, {Ramirez}, {Sab{\'{\i}}n-Sanjuli{\'a}n}, {Smartt},
  {Stroud}, {van Loon}, {Vink}, \& {Walborn}}]{Evans_2011}
{Evans}, C.~J., {Taylor}, W.~D., {H{\'e}nault-Brunet}, V., {et~al.} 2011, \aap,
  530, A108+

\bibitem[{{Fitzpatrick} \& {Garmany}(1990)}]{Fitzpatrick_1990}
{Fitzpatrick}, E.~L. \& {Garmany}, C.~D. 1990, \apj, 363, 119

\bibitem[{{Fullerton} {et~al.}(2006){Fullerton}, {Massa}, \&
  {Prinja}}]{Fullerton_2006}
{Fullerton}, A.~W., {Massa}, D.~L., \& {Prinja}, R.~K. 2006, \apj, 637, 1025

\bibitem[{{Goldreich} \& {Kumar}(1990)}]{Goldreich_1990}
{Goldreich}, P. \& {Kumar}, P. 1990, \apj, 363, 694

\bibitem[{{Gr{\"a}fener} \& {Hamann}(2008)}]{Graefener_2008}
{Gr{\"a}fener}, G. \& {Hamann}, W.-R. 2008, \aap, 482, 945

\bibitem[{{Gr{\"a}fener} {et~al.}(2012){Gr{\"a}fener}, {Owocki}, \&
  {Vink}}]{Graefener_2012}
{Gr{\"a}fener}, G., {Owocki}, S.~P., \& {Vink}, J.~S. 2012, \aap, 538, A40

\bibitem[{{Gr{\"a}fener} \& {Vink}(2013)}]{Graefener_2013}
{Gr{\"a}fener}, G. \& {Vink}, J.~S. 2013, \aap, 560, A6

\bibitem[{{Gr{\"a}fener} {et~al.}(2011){Gr{\"a}fener}, {Vink}, {de Koter}, \&
  {Langer}}]{Graefener_2011}
{Gr{\"a}fener}, G., {Vink}, J.~S., {de Koter}, A., \& {Langer}, N. 2011, \aap,
  535, A56

\bibitem[{{Hamann} {et~al.}(1995){Hamann}, {Koesterke}, \&
  {Wessolowski}}]{Hamann_1995}
{Hamann}, W.-R., {Koesterke}, L., \& {Wessolowski}, U. 1995, \aap, 299, 151

\bibitem[{{Heger} \& {Langer}(1996)}]{Heger_1996}
{Heger}, A. \& {Langer}, N. 1996, \aap, 315, 421

\bibitem[{{Heger} \& {Langer}(2000)}]{Heger_2000b}
{Heger}, A. \& {Langer}, N. 2000, \apj, 544, 1016

\bibitem[{{Heger} {et~al.}(2000){Heger}, {Langer}, \& {Woosley}}]{Heger_2000}
{Heger}, A., {Langer}, N., \& {Woosley}, S.~E. 2000, \apj, 528, 368

\bibitem[{{Humphreys} \& {Davidson}(1994)}]{Humphreys_1994}
{Humphreys}, R.~M. \& {Davidson}, K. 1994, \pasp, 106, 1025

\bibitem[{{Hunter} {et~al.}(2008){Hunter}, {Brott}, {Lennon}, {Langer},
  {Dufton}, {Trundle}, {Smartt}, {de Koter}, {Evans}, \& {Ryans}}]{Hunter_2008}
{Hunter}, I., {Brott}, I., {Lennon}, D.~J., {et~al.} 2008, \apjl, 676, L29

\bibitem[{{Ishii} {et~al.}(1999){Ishii}, {Ueno}, \& {Kato}}]{Ishii_1999}
{Ishii}, M., {Ueno}, M., \& {Kato}, M. 1999, \pasj, 51, 417

\bibitem[{{Joss} {et~al.}(1973){Joss}, {Salpeter}, \& {Ostriker}}]{Joss_1973}
{Joss}, P.~C., {Salpeter}, E.~E., \& {Ostriker}, J.~P. 1973, \apj, 181, 429

\bibitem[{{Kato}(1986)}]{Kato_1986}
{Kato}, M. 1986, \apss, 119, 57

\bibitem[{{Kippenhahn} \& {Weigert}(1990)}]{Kippenhahn_1990}
{Kippenhahn}, R. \& {Weigert}, A. 1990, {Stellar Structure and Evolution}

\bibitem[{{K{\"o}hler} {et~al.}(2012){K{\"o}hler}, {Borzyszkowski}, {Brott},
  {Langer}, \& {de Koter}}]{Koehler_2012}
{K{\"o}hler}, K., {Borzyszkowski}, M., {Brott}, I., {Langer}, N., \& {de
  Koter}, A. 2012, \aap, 544, A76

\bibitem[{{Kozyreva} {et~al.}(2014){Kozyreva}, {Blinnikov}, {Langer}, \&
  {Yoon}}]{Kozyreva_2014}
{Kozyreva}, A., {Blinnikov}, S., {Langer}, N., \& {Yoon}, S.-C. 2014, \aap,
  565, A70

\bibitem[{{Langer}(1989)}]{Langer_1989}
{Langer}, N. 1989, \aap, 210, 93

\bibitem[{{Langer}(1991)}]{Langer_1991}
{Langer}, N. 1991, \aap, 252, 669

\bibitem[{{Langer}(1992)}]{Langer_1992}
{Langer}, N. 1992, \aap, 265, L17

\bibitem[{{Langer}(1997)}]{Langer_1997}
{Langer}, N. 1997, in Astronomical Society of the Pacific Conference Series,
  Vol. 120, Luminous Blue Variables: Massive Stars in Transition, ed. A.~{Nota}
  \& H.~{Lamers}, 83

\bibitem[{{Langer}(1998)}]{Langer_1998}
{Langer}, N. 1998, \aap, 329, 551

\bibitem[{{Langer}(2012)}]{Langer_2012}
{Langer}, N. 2012, \araa, 50, 107

\bibitem[{{Langer} {et~al.}(1983){Langer}, {Fricke}, \&
  {Sugimoto}}]{Langer_1983}
{Langer}, N., {Fricke}, K.~J., \& {Sugimoto}, D. 1983, \aap, 126, 207

\bibitem[{{Langer} \& {Kudritzki}(2014)}]{Langer_2014}
{Langer}, N. \& {Kudritzki}, R.~P. 2014, \aap, 564, A52

\bibitem[{{Langer} {et~al.}(2007){Langer}, {Norman}, {de Koter}, {Vink},
  {Cantiello}, \& {Yoon}}]{Langer_2007}
{Langer}, N., {Norman}, C.~A., {de Koter}, A., {et~al.} 2007, \aap, 475, L19

\bibitem[{{Lucy} \& {Abbott}(1993)}]{Lucy_1993}
{Lucy}, L.~B. \& {Abbott}, D.~C. 1993, \apj, 405, 738

\bibitem[{{Maeder} \& {Meynet}(2000)}]{Maeder_2000}
{Maeder}, A. \& {Meynet}, G. 2000, \araa, 38, 143

\bibitem[{{Maeder} \& {Meynet}(2011)}]{Maeder_2011}
{Maeder}, A. \& {Meynet}, G. 2011, ArXiv e-prints

\bibitem[{{McWilliam} \& {Rauch}(2004)}]{McWilliam_2004}
{McWilliam}, A. \& {Rauch}, M. 2004, Origin and Evolution of the Elements

\bibitem[{{Mokiem} {et~al.}(2007){Mokiem}, {de Koter}, {Evans}, {Puls},
  {Smartt}, {Crowther}, {Herrero}, {Langer}, {Lennon}, {Najarro}, {Villamariz},
  \& {Vink}}]{Mokiem_2007}
{Mokiem}, M.~R., {de Koter}, A., {Evans}, C.~J., {et~al.} 2007, \aap, 465, 1003

\bibitem[{{Muijres} {et~al.}(2012){Muijres}, {Vink}, {de Koter}, {M{\"u}ller},
  \& {Langer}}]{Muijres_2012}
{Muijres}, L.~E., {Vink}, J.~S., {de Koter}, A., {M{\"u}ller}, P.~E., \&
  {Langer}, N. 2012, \aap, 537, A37

\bibitem[{{M{\"u}ller} \& {Vink}(2008)}]{Mueller_2008}
{M{\"u}ller}, P.~E. \& {Vink}, J.~S. 2008, \aap, 492, 493

\bibitem[{{M{\"u}ller} \& {Vink}(2014)}]{Mueller_2014}
{M{\"u}ller}, P.~E. \& {Vink}, J.~S. 2014, \aap, 564, A57

\bibitem[{{Nieuwenhuijzen} \& {de Jager}(1990)}]{Nieuwenhuijzen_1990}
{Nieuwenhuijzen}, H. \& {de Jager}, C. 1990, \aap, 231, 134

\bibitem[{{Nugis} \& {Lamers}(2000)}]{Nugis_2000}
{Nugis}, T. \& {Lamers}, H.~J.~G.~L.~M. 2000, \aap, 360, 227

\bibitem[{{Petrovic} {et~al.}(2005){Petrovic}, {Langer}, {Yoon}, \&
  {Heger}}]{Petrovic_2005}
{Petrovic}, J., {Langer}, N., {Yoon}, S.-C., \& {Heger}, A. 2005, \aap, 435,
  247

\bibitem[{{Petrovic} {et~al.}(2006){Petrovic}, {Pols}, \&
  {Langer}}]{Petrovic_2006}
{Petrovic}, J., {Pols}, O., \& {Langer}, N. 2006, \aap, 450, 219

\bibitem[{{Ram{\'{\i}}rez-Agudelo} {et~al.}(2013){Ram{\'{\i}}rez-Agudelo},
  {Sim{\'o}n-D{\'{\i}}az}, {Sana}, {de Koter}, {Sab{\'{\i}}n-Sanjul{\'{\i}}an},
  {de Mink}, {Dufton}, {Gr{\"a}fener}, {Evans}, {Herrero}, {Langer}, {Lennon},
  {Ma{\'{\i}}z Apell{\'a}niz}, {Markova}, {Najarro}, {Puls}, {Taylor}, \&
  {Vink}}]{Agudelo_2013}
{Ram{\'{\i}}rez-Agudelo}, O.~H., {Sim{\'o}n-D{\'{\i}}az}, S., {Sana}, H.,
  {et~al.} 2013, \aap, 560, A29

\bibitem[{{Sana} {et~al.}(2013{\natexlab{a}}){Sana}, {de Koter}, {de Mink},
  {Dunstall}, {Evans}, {H{\'e}nault-Brunet}, {Ma{\'{\i}}z Apell{\'a}niz},
  {Ram{\'{\i}}rez-Agudelo}, {Taylor}, {Walborn}, {Clark}, {Crowther},
  {Herrero}, {Gieles}, {Langer}, {Lennon}, \& {Vink}}]{Sana_2013b}
{Sana}, H., {de Koter}, A., {de Mink}, S.~E., {et~al.} 2013{\natexlab{a}},
  \aap, 550, A107

\bibitem[{{Sana} {et~al.}(2012){Sana}, {de Mink}, {de Koter}, {Langer},
  {Evans}, {Gieles}, {Gosset}, {Izzard}, {Le Bouquin}, \&
  {Schneider}}]{Sana_2012}
{Sana}, H., {de Mink}, S.~E., {de Koter}, A., {et~al.} 2012, Science, 337, 444

\bibitem[{{Sana} {et~al.}(2013{\natexlab{b}}){Sana}, {van Boeckel}, {Tramper},
  {Ellerbroek}, {de Koter}, {Kaper}, {Moffat}, {Schnurr}, {Schneider}, \&
  {Gies}}]{Sana_2013}
{Sana}, H., {van Boeckel}, T., {Tramper}, F., {et~al.} 2013{\natexlab{b}},
  \mnras, 432, L26

\bibitem[{{Schneider} {et~al.}(2014){Schneider}, {Izzard}, {de Mink}, {Langer},
  {Stolte}, {de Koter}, {Gvaramadze}, {Hu{\ss}mann}, {Liermann}, \&
  {Sana}}]{Schneider_2014}
{Schneider}, F.~R.~N., {Izzard}, R.~G., {de Mink}, S.~E., {et~al.} 2014, \apj,
  780, 117

\bibitem[{{Schnurr} {et~al.}(2008){Schnurr}, {Casoli}, {Chen{\'e}}, {Moffat},
  \& {St-Louis}}]{Schnurr_2008}
{Schnurr}, O., {Casoli}, J., {Chen{\'e}}, A.-N., {Moffat}, A.~F.~J., \&
  {St-Louis}, N. 2008, \mnras, 389, L38

\bibitem[{{Schnurr} {et~al.}(2009){Schnurr}, {Moffat}, {Villar-Sbaffi},
  {St-Louis}, \& {Morrell}}]{Schnurr_2009}
{Schnurr}, O., {Moffat}, A.~F.~J., {Villar-Sbaffi}, A., {St-Louis}, N., \&
  {Morrell}, N.~I. 2009, \mnras, 395, 823

\bibitem[{{Smith}(2013)}]{Smith_2013}
{Smith}, N. 2013, \mnras, 429, 2366

\bibitem[{{Spruit}(2002)}]{Spruit_2002}
{Spruit}, H.~C. 2002, \aap, 381, 923

\bibitem[{{Spruit}(2006)}]{Spruit_2006}
{Spruit}, H.~C. 2006, ArXiv Astrophysics e-prints

\bibitem[{{Vink} {et~al.}(2010){Vink}, {Brott}, {Gr{\"a}fener}, {Langer}, {de
  Koter}, \& {Lennon}}]{Vink_2010}
{Vink}, J.~S., {Brott}, I., {Gr{\"a}fener}, G., {et~al.} 2010, \aap, 512, L7+

\bibitem[{{Vink} {et~al.}(1999){Vink}, {de Koter}, \& {Lamers}}]{Vink_1999}
{Vink}, J.~S., {de Koter}, A., \& {Lamers}, H.~J.~G.~L.~M. 1999, \aap, 350, 181

\bibitem[{{Vink} {et~al.}(2000){Vink}, {de Koter}, \& {Lamers}}]{Vink_2000}
{Vink}, J.~S., {de Koter}, A., \& {Lamers}, H.~J.~G.~L.~M. 2000, \aap, 362, 295

\bibitem[{{Vink} {et~al.}(2001){Vink}, {de Koter}, \& {Lamers}}]{Vink_2001}
{Vink}, J.~S., {de Koter}, A., \& {Lamers}, H.~J.~G.~L.~M. 2001, \aap, 369, 574

\bibitem[{{Vink} \& {Gr{\"a}fener}(2012)}]{Vink_2012}
{Vink}, J.~S. \& {Gr{\"a}fener}, G. 2012, \apjl, 751, L34

\bibitem[{{Vink} {et~al.}(2011){Vink}, {Muijres}, {Anthonisse}, {de Koter},
  {Gr{\"a}fener}, \& {Langer}}]{Vink_2011}
{Vink}, J.~S., {Muijres}, L.~E., {Anthonisse}, B., {et~al.} 2011, \aap, 531,
  A132

\bibitem[{{Woosley} \& {Heger}(2006)}]{Woosley_2006}
{Woosley}, S.~E. \& {Heger}, A. 2006, \apj, 637, 914

\bibitem[{{Yoon} \& {Langer}(2005)}]{Yoon_2005}
{Yoon}, S. \& {Langer}, N. 2005, \aap, 443, 643

\bibitem[{{Yoon} {et~al.}(2012){Yoon}, {Dierks}, \& {Langer}}]{Yoon_2012}
{Yoon}, S.-C., {Dierks}, A., \& {Langer}, N. 2012, \aap, 542, A113

\bibitem[{{Yoon} {et~al.}(2010){Yoon}, {Woosley}, \& {Langer}}]{Yoon_2010}
{Yoon}, S.-C., {Woosley}, S.~E., \& {Langer}, N. 2010, \apj, 725, 940

\bibitem[{{Yusof} {et~al.}(2013){Yusof}, {Hirschi}, {Meynet}, {Crowther},
  {Ekstr{\"o}m}, {Frischknecht}, {Georgy}, {Abu Kassim}, \&
  {Schnurr}}]{Yusof_2013}
{Yusof}, N., {Hirschi}, R., {Meynet}, G., {et~al.} 2013, \mnras, 433, 1114

\end{thebibliography}

\end{document}